\documentclass{sig-alternate} 
\pdfoutput=1
\usepackage{mathptmx} 
\newcommand{\ignore}[1]{}
\usepackage{fancyhdr}
\usepackage[normalem]{ulem}
\usepackage{microtype}
\usepackage{caption}
\usepackage{subcaption}

\DeclareMathAlphabet{\mathcal}{OMS}{cmsy}{m}{n}

\usepackage{amsmath, amssymb}
\usepackage{algorithm}
\usepackage[noend]{algpseudocode}
\usepackage[mathscr]{euscript}
\usepackage[dvipsnames]{xcolor}
\usepackage{mathtools}
\usepackage{appendix}
\usepackage{multirow}
\usepackage{paralist}
\usepackage[bookmarks=true,breaklinks=true,letterpaper=true,colorlinks,linkcolor=black,citecolor=blue,urlcolor=black]{hyperref}
\usepackage{authblk}

\usepackage{tabularx}

\usepackage{listings}
\usepackage{courier}
\usepackage{tikz}
\usepackage{xcolor}
\newcommand*\circled[1]{\tikz[baseline=(char.base)]{
    \node[shape=circle,fill,inner sep=1pt,scale=0.8] (char) {\textcolor{white}{#1}};}}
\newcommand*\wcircled[1]{\tikz[baseline=(char.base)]{
    \node[shape=circle,draw=black,inner sep=1pt,scale=0.8] (char) {\textcolor{black}{#1}};}}
\usepackage{fixltx2e}
\usepackage{enumitem}

\title{Continual Learning Approach for Improving the Data and Computation Mapping in Near-Memory Processing System}
\author[1]{Pritam Majumder}
\author[2]{Jiayi Huang}
\author[1]{Sungkeun Kim}
\author[1]{Abdullah Muzahid}
\author[1]{Dylan Siegers}
\author[1]{Chia-Che Tsai}
\author[1]{Eun Jung Kim}
\affil[1]{(pritam2309, ksungkeun84, Abdullah.Muzahid, dsiegers, chiache, ejkim)@tamu.edu, Texas A\&M University}
\affil[2]{jyhuang@ucsb.edu, UC Santa Barbara}

\newcommand{\rl}{{RL}}
\newcommand{\tech}{\texttt{AIMM}}

\begin{document}
\maketitle
\pagestyle{plain}

\section{abstract}
The resurgence of near-memory processing (NMP) with the advent of big data has shifted the computation paradigm from processor-centric to memory-centric computing. To meet the bandwidth and capacity demands of memory-centric computing, 3D memory has been adopted to form a scalable memory-cube network.
Along with NMP and memory system development, the  mapping for placing data and guiding computation in the memory-cube network has become crucial in driving the performance improvement in NMP. However, it is very challenging to design a universal optimal mapping for all applications due to unique application behavior and intractable decision space.

In this paper, we propose an artificially intelligent memory mapping scheme, \tech, that optimizes data placement and resource utilization through page and computation remapping.  Our proposed technique involves continuously evaluating and learning the impact of mapping decisions on system performance for any application. \tech\ uses a neural network to achieve a near-optimal mapping during execution, trained using a reinforcement learning algorithm that is known to be effective for exploring a vast design space. We also provide a detailed \tech\ hardware design that can be adopted as a plugin module for various NMP systems. Our experimental evaluation shows that \tech\ improves the baseline NMP performance in single and multiple program scenario by up to 70\% and 50\%, respectively.


\section{Introduction}
\label{intro}
With the explosion of data, emerging applications such as machine learning and graph processing~\cite{thomas2014cortexsuite, ahmad2015crono} have driven copious data movement across the modern memory hierarchy. Due to the limited bandwidth of traditional double data rate (DDR) memory interfaces, memory wall becomes a major bottleneck for system performance.
Consequently, 3D-stacked memory cubes such as the hybrid memory cube (HMC)~\cite{pawlowski2011hmc} and high bandwidth memory (HBM)~\cite{lee2014hbm} were invented to provide high bandwidth that suffices the tremendous bandwidth requirement of big data applications.
Although high-bandwidth stacked memory has reduced the bandwidth pressure, modern processor-centric computation fails to process large data sets efficiently due to the expensive data movement of low-reuse data.
To avoid the high cost of data movement, near-memory processing (NMP) was revived and enabled memory-centric computing by moving the computation close to the data in the memory~\cite{ahn2016pei,ahn2015tesseract,nai2017graphpim,huang2019active,hsieh2016tom}.

Recently, memory-centric NMP systems demand even larger memory capacity to accommodate the increasing sizes of data sets and workloads.
For example, the recent GPT-3 transformer model~\cite{brown2020gpt3} has 175 billion parameters and requires at least 350 GB memory to load a model for inference and even more memory for training.
As a solution, multiple memory cubes can be combined to satisfy the high demands~\cite{hsieh2016tom}.
For instance, the recently announced AMD Radeon VII and NVIDIA A100 GPU have each included 4 and 6 HBMs, respectively.
More memory cubes are adopted in recent works to form a memory-cube network~\cite{kim2013mcn,zhan2016umemnet} for further scale-up.
Moreover, NMP support for memory-cube network has also been investigated in recent proposals to accelerate data-intensive applications~\cite{huang2019active}.

\begin{figure*}[t]
\centering
    \begin{subfigure}{0.33\textwidth}
    \includegraphics[width=0.9\textwidth]{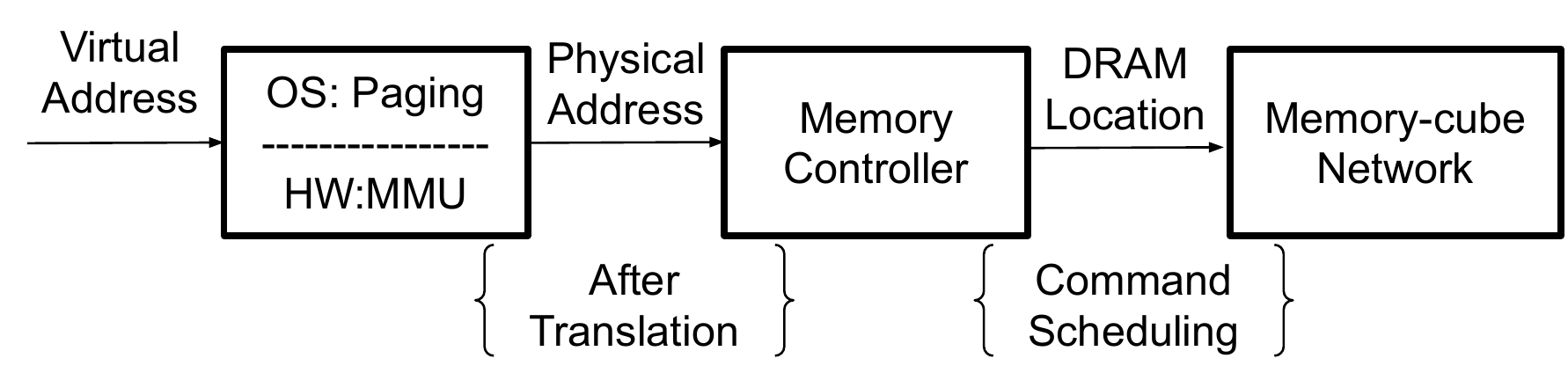}
    \caption{Data mapping in conventional system}
    \label{fig:data_mapping}
    \end{subfigure}
    \begin{subfigure}{0.33\textwidth}
    \includegraphics[scale=0.3]{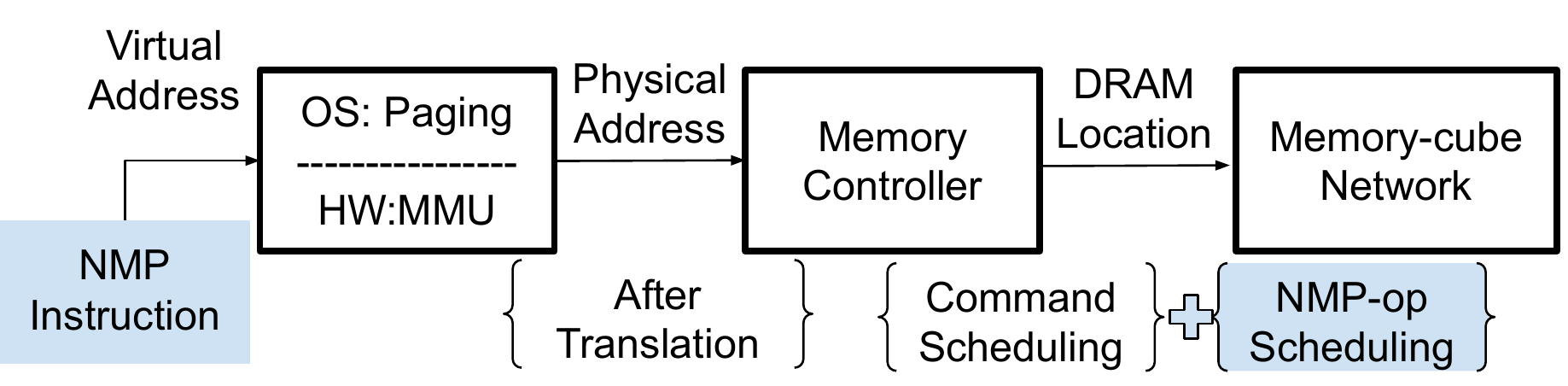}
    \caption{Memory (data/compute) mapping in NMP system}
    \label{fig:nmp_mapping}
    \end{subfigure}
    \begin{subfigure}{0.33\textwidth}
    \includegraphics[scale=0.3]{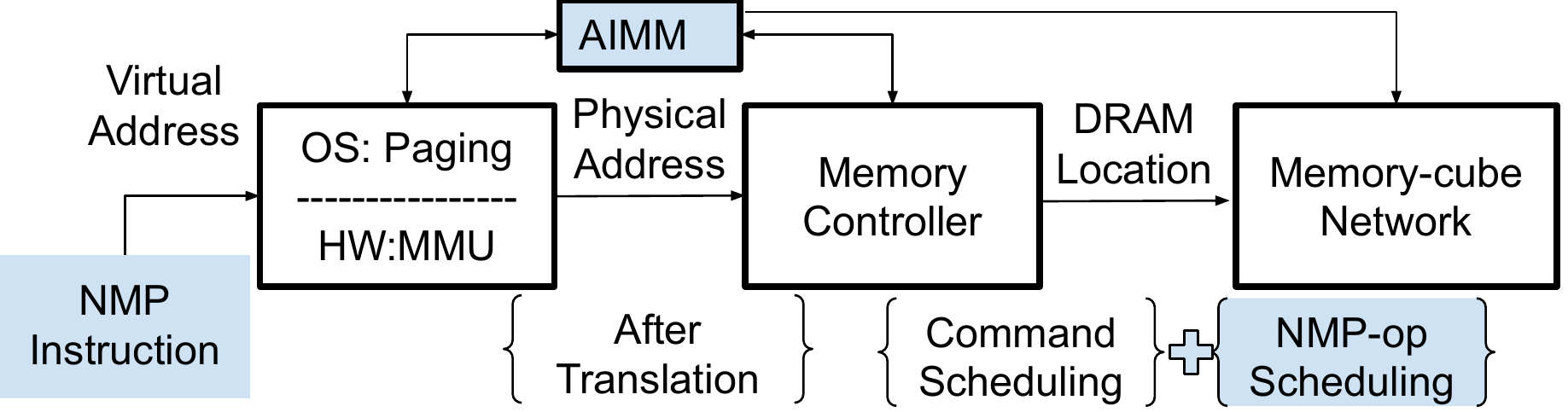}
    \caption{Memory mapping in NMP system with \tech}
    \label{fig:aimm_mapping}
    \end{subfigure}
    \vspace{-0.2cm}
    \caption{Memory mapping in different systems: (a) data mapping in conventional system, (b) data and compute mapping in NMP system, and (c) data and compute mapping in NMP system with \tech{}.}
    \vspace{-0.2cm}
    \label{fig:mapping}
\end{figure*}

Along with the NMP and memory system developments, memory mapping for placing data and computation in the memory-cube network has become an important design consideration for NMP system performance. 
Figure~\ref{fig:data_mapping} shows an overview of the data mapping in two steps---virtual-to-physical page mapping and physical-to-DRAM address mapping. 
The paging system maps a virtual page to a physical page frame, then the memory controller hashes the physical address to DRAM address, which identifies the location in the DRAM.
For NMP systems, the memory mapping should handle computation as well as data, as shown in Figure~\ref{fig:nmp_mapping}.
Besides the data mapping, the memory controller also decides the memory cube in which the NMP operation is to be scheduled.

Many works have focused on physical-to-DRAM address mapping to improve the memory-level parallelism~\cite{zhang2000permutation,akin2015data,liu2018get}.
Recently, DRAM address mapping has been investigated in NMP systems to better co-locate data in the same cubes~\cite{hsieh2016tom}.
However, adapting the mapping for the dynamic NMP application behavior is problematic due to the possibility of excessive data migration for every memory byte in order to reflect the new mappings.
On the other hand, virtual-to-physical page frame mapping provides an alternative approach to adjust the data mapping during run time. 
Although such research has existed for processor-centric NUMA systems~\cite{piccoli14migration, goglin09migration, chiang18numa, hinuma}, it has not yet been explored for the memory-centric NMP systems,
where computation is finer grained and tightly coupled with data in the memory system.

Furthermore, recent NMP research~\cite{ahn2016pei,huang2019active} tightly couple the computation mapping with given data mapping for static offloading, which has not considered the co-exploration of the intermingling between them. This may cause computation resource under utilization in different cubes and lead to performance degradation, especially for irregular data such as graphs and sparse matrices. Thus, it is challenging to achieve an optimal memory mapping for NMP with another dimension of computation mapping. Moreover, the unique and dynamic application behaviors as well as the intractable decision space (more than $10^{14}$) with NMP in the memory-cube network make it even more challenging to design a universal optimal mapping for all types of workloads.

In this paper, we propose an artificially intelligent memory mapping scheme, \tech\,, that optimizes data placement and resource utilization through page and computation remapping. \tech{} can adapt to different applications and runtime behaviors by continuously evaluating and learning the impact of mapping decisions on system performance. It uses a neural network to learn the near-optimal mapping at any instance during execution, and is trained using reinforcement learning which is known to excel at exploring vast design space. Figure~\ref{fig:aimm_mapping} depicts a system overview augmented by \tech{}. In the proposed system, \tech{} interacts with the paging system, the memory controller, and the memory-cube network.
It continuously evaluates the NMP performance through the memory controller and makes data remapping decisions through the paging and page migration systems in the memory network. It is also consulted for computation remapping to improve NMP operation scheduling and performance.

The contributions of this paper are as follows.

\begin{compactitem}
    \item An artificially intelligent memory mapping, \tech{}, for adjusting the memory mapping dynamically to adapt to application behavior and to improve resource utilization in NMP systems. To the best of our knowledge, this is the first NMP work that targets mapping of both data and computation.
    \item A reinforcement learning formulation to explore the vast design space and to train a neural network agent for the memory mapping problem.
    \item A detailed hardware design and practical implementation in a plug-and-play manner to be applied in various NMP systems.
    \item A comprehensive experimental results which show that the proposed \tech{} improves the performance of state-of-the-art NMP systems in single and multi program scenario by up to 70\% and 50\% respectively.
\end{compactitem}

The rest of the paper is organized as follows.
In \S\ref{background} we discuss the background and related work.
Next, we state the problem and formulate the proposed approach as a reinforcement learning problem in \S\ref{formulation}. In \S\ref{hardware_implementation}, we propose a practical hardware design for \tech{} followed by the evaluation methodology in \S\ref{sec:methodology}. Then, we present and analyze the results in \S\ref{results}.` Finally, we conclude this work in \S\ref{conclusion}.

\section{Background and Related Work}
\label{background}

In this section, we introduce the background and related works, including near-memory processing, memory mapping and management, as well as reinforcement learning and its architecture applications.

\subsection{Near-Memory Processing}

Recently, near-memory processing (NMP)~\cite{ahn2016pei,ahn2015tesseract,hsieh2016tom,nai2017graphpim,huang2019active} was revived with the advent of 3D-stacked memory~\cite{pawlowski2011hmc,lee2014hbm}.
In an NMP system with 3D memory cubes, the processing capability is in the base logic die under a stack of DRAM layers to utilize the ample internal bandwidth~\cite{ahn2016pei}.
Later research also proposes near-bank processing with logic near memory banks in the same DRAM layer to exploit even higher bandwidth~\cite{yazdanbakhsh2018dram,gu2020ipim}, such as FIMDRAM~\cite{kwon202125} announced recently by Samsung.
Recent proposals~\cite{alian2019netdimm,alian2018application,huangfu2019medal,kwon2019tensordimm,ke2020recnmp} have also explored augmenting traditional DIMMs with computation in the buffer die to provide low-cost but bandwidth-limited NMP solutions.

For computation offloading, there are mainly two ways: instruction offloading and kernel offloading.
PIM-Enabled Instruction (PEI)~\cite{ahn2016pei}, for instance, offloads instructions from the CPU to memory using the extended ISA.
Other works such as GraphPim~\cite{nai2017graphpim}, Active-Routing~\cite{huang2019active}, FIMDRAM~\cite{kwon202125}, and DIMM NMP solutions~\cite{alian2019netdimm, huangfu2019medal, kwon2019tensordimm, ke2020recnmp} also fall in the same category.
For kernel offloading, similar to GPU, kernel code blocks are offloaded to the memory through library and low-level drivers, such as Tesseract~\cite{ahn2015tesseract}, TOM~\cite{hsieh2016tom}.

In this work, we study the memory mapping problem and propose \tech{} for data and computation placement for NMP systems.
We use physical-to-DRAM address mapping to improve data co-location and to reduce off-chip data movement in recent research~\cite{hsieh2016tom}.
However, we cannot adjust DRAM mapping during application runs to adapt to application behaviors,
as such adjustment
would trigger migration for every memory byte and incur prohibitive performance overhead.
On the other hand, we explore virtual-to-physical mapping through page migration to achieve low-cost dynamic remapping.
To explore the missed opportunity of dynamic 
computation remapping, 
\tech{} co-explores data and computation mapping holistically in NMP to further drive the performance improvement.
We showcase with several NMP solutions along with processing-in-logic layers and instruction offloading to demonstrate the effectiveness of \tech{}.
Since our solution is decoupled from the NMP design, it is applicable to various NMP alternatives.

\subsection{Heuristic-based Memory Management}

Traditionally, operating systems
(including kernels, system libraries, and runtimes) employ a heuristic approach~\cite{tanenbaum-osbook}, opting for solutions which promote locality~\cite{hoard, locality_of_ref, dynamic-storage-allocation, locality-improving_DMA} and defragmentation~\cite{dynamic-storage-allocation, buddy} for more efficient memory management.
An OS's memory management system typically includes page and object allocators (either in-applications or in-kernel)~\cite{hoard, dynamic-storage-allocation, buddy}, kernel page fault handlers~\cite{lkml-transparent-huge-pages, hawkeye}, and application garbage collectors~\cite{locality_of_ref, lisp-gc, lieberman-gc-83}.
These components either collaboratively or competitively manage a pool or pools of virtual or physical memory, based on demands from users and limited system views made available to the components.
As a result, optimizations for memory management in existing OSes are oftentimes best-effort, without exact certainty that the performance of the system or an application will be improved under the deployed solutions.

Take locality for example, an OS's memory management system generally aims for improving the spatial or temporal locality in software,
in order to better utilize the parallelism of hardware and/or to reduce the cost of data movement.
A thread-private allocator like HOARD~\cite{hoard} uses bulk allocation and per-thread free lists
to try to ensure the proximity of memory access within one thread. 
A generational garbage collector~\cite{generation-gc, java-gc} exploits the difference between short-lived and long-lived objects,
to rally the most frequent memory allocation and reclamation to a relatively small heap which fits into the last-level cache.
For Non-Uniform Memory Access (NUMA)~\cite{bolosky1991numa} architecture,
Linux takes a zone-based approach to assign physical memory to processes based on CPU affinity,
as well as migrates physical pages actively
to localize memory accesses in CPU nodes.

Page migration~\cite{piccoli14migration, nikolopoulos00migration}, in particular, is a technique used in different systems to balance memory and CPU utilization and to avoid high memory latency caused by distance.
Page migration is proven useful in a NUMA system, or a heterogeneous system such a CPU-GPU or multi-GPU shared memory system.
In such a shared memory system, a GPU that frequently accesses far memory can lead to high performance overhead.
A ``first-touch'' approach~\cite{yao81mem-alloc, ribeiro09affinity, lameter13numa} can localize memory allocation to the first thread or the first GPU who accesses the memory, but does not guarantee the locality in future computation.
Page migration can either be triggered by an OS which monitors the system performance,
or be requested by the applications
based on the knowledge of application logic.
Baruah et al.~\cite{baruah2020griffin} proposes a dynamic page classification mechanism to select pages and migrate them to the GPU that is predicted to access the pages most frequently in the future.
In this work, we co-explore both data movement and computation scheduling to improve resource utilization. 

%

\begin{figure}[t]
    \centering
    \includegraphics[scale=0.5]{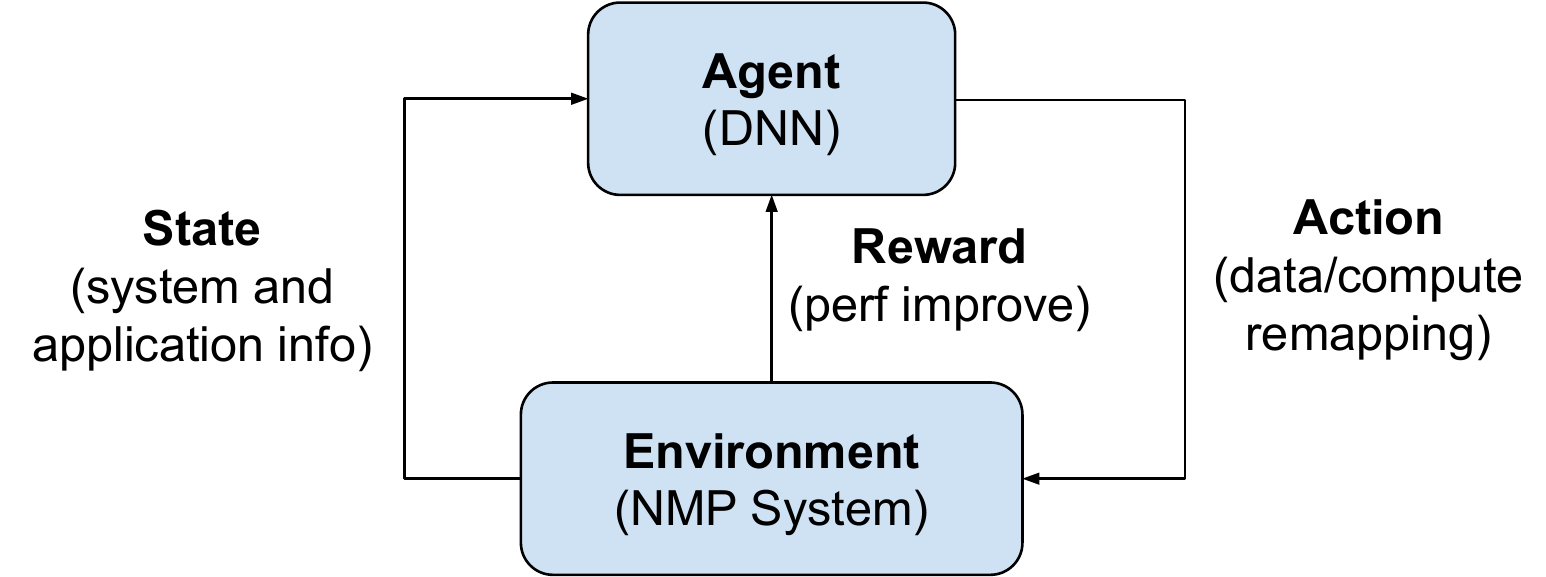}
    \caption{A typical reinforcement learning framework.}
    \label{fig:rl}
\end{figure}

\subsection{Reinforcement Learning (RL)}

RL is a machine learning method where an agent explores actions in an environment to maximize the accumulative rewards~\cite{sutton2018reinforcement}.
In this paper, RL is used to design an agent that explores data and computation remapping decisions (actions) in the NMP system (environment) to maximize the performance (rewards).
Figure~\ref{fig:rl} shows a typical setting of RL, where an agent interacts with the environment {$\mathcal{E}$} over a number of discrete time steps.
At each time step $t$, the agent observes an environment state $s_t \in \mathcal{S}$ \textit{(state space)}, and selects an action $a_t \in \mathcal{A}$ \textit{(action space)} to change the environment according to its policy $\pi$, which is a mapping from $s_t$ to $a_t$.
In return, it receives a reward $r_t$ and a new state $s_{t+1}$ and continues the same process.
The accumulative rewards after time step $t$ can be calculated as
\begin{equation}
R_t = \sum{_{t'=t}^T} \gamma{^{t'-t}r_{t'}}
\end{equation}
where $\gamma \in (0,1]$ is the discount factor and $T$ is the time step when the process terminates.
The state-action value function $Q_\pi(s,a)$ under policy $\pi$ can be expressed as
\begin{equation}
    Q_\pi(s,a) = \mathbb{E}_\pi[R_t|s_t = s, a_t = a]
\end{equation}
The objective of the agent is to maximize the expected reward.

In this paper, we use Q-learning~\cite{mnih2015human} that selects the optimal action $a^*$ that obtains the highest state-action value (Q value) $Q^*_\pi(s,a^*) =$ max $\mathbb{E}_\pi[R_t|s_t = s, a_t = a^*]$ following the \textit{Bellman equation}~\cite{sutton2018reinforcement}.
The Q value is typically estimated by a function approximator.
In this work, we use a deep neural network with parameters $\theta$ as the function approximator to predict the optimal Q value $Q(s,a;\theta) \approx Q^*(s,a)$.
Given the predicted value $Q(s_t,a_t;\theta)$, the neural network is trained by minimizing the squared loss:
\begin{equation}
    L(\theta) = (y - Q(s_t,a_t;\theta))^2,
\end{equation}
where $y = r_t(s_t, a_t) + \gamma$ max\textsubscript{$a'$} $Q(s_{t+1}, a';\theta|s_t,a_t)$ is the target that represents the highest accumulative rewards of the next state $s_{t+1}$ if action $a_t$ is chosen for state $s_t$.

RL has been widely applied to network-on-chip (NoC) for power management~\cite{won2014up}, loop placement for routerless NoC~\cite{lin2020deep}, arbitration policy~\cite{yin2020experiences}, and reliability~\cite{wang2019intellinoc}.
It has also been used for designing memory systems, such as prefetching~\cite{peled2015semantic} and memory controller~\cite{ipek2008self}.
Additionally, it has been applied to DNN compilation and mapping optimization~\cite{ahn2019reinforcement,kao2020confuciux,wu2020core}.
In this work, we use RL for co-exploration of data and computation mapping in NMP systems.

\section{Proposed Approach}
\label{formulation}

In this section, we introduce our proposed approach by first presenting the overview and problem formulation.
Then we describe the representations of state, action and reward function for \tech{} NMP memory mapping, followed by the DNN architecture and training of the RL-based agent.


\subsection{Overview and Problem Formulation}

\begin{figure*}
    \centering
    \includegraphics[scale=0.37]{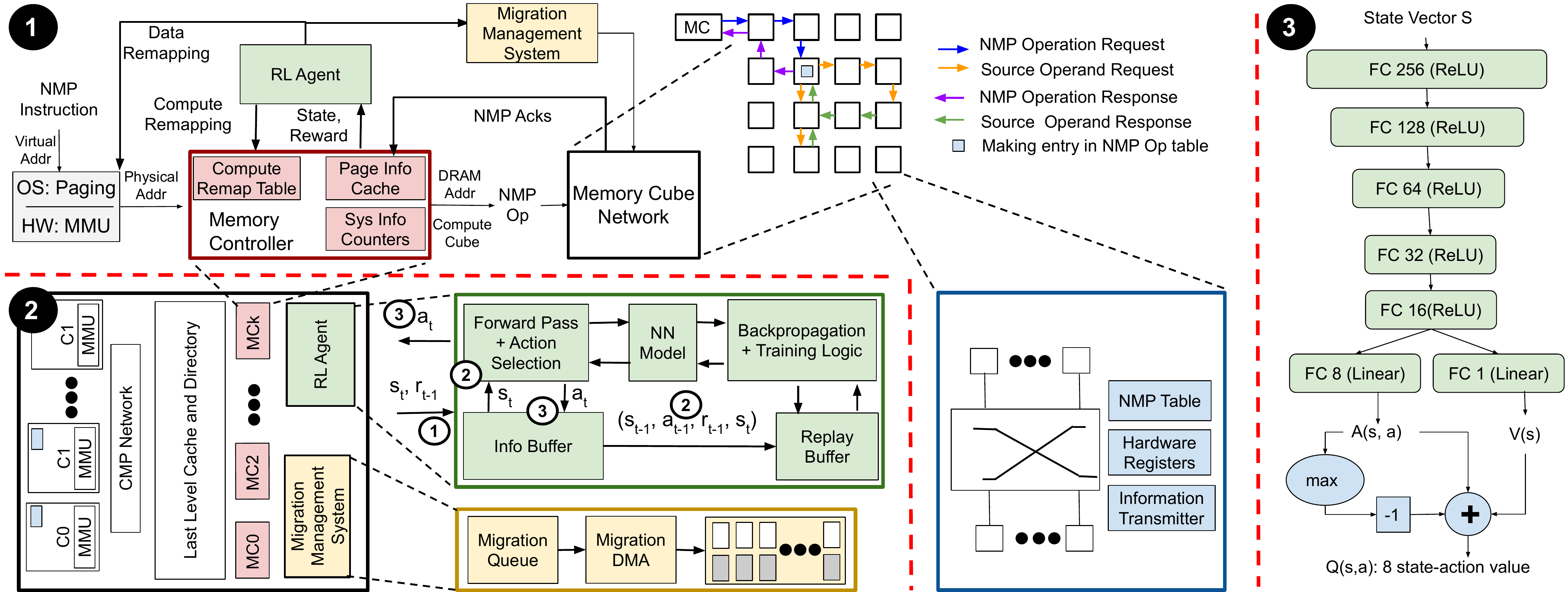}
    \caption{\tech{} reinforcement learning framework and system architecture: (1) overview of \tech{} memory mapping for data and computation in NMP systems, (2) system architecture, and (3) Dueling network for RL-based agent.}
    \label{fig:sys_flow_arch}
\end{figure*}

Figure~\ref{fig:sys_flow_arch} \circled{1} depicts the overview of the proposed 
\tech{}, which seeks to optimize data placement and computation distribution through page and computation remapping\footnote{For the first time seen pages and their associated NMP operations, the default data mapping and default compute cube are decided by the OS and the memory controller, respectively.}.
In short, the RL agent keeps interacting with the environment, which includes the paging system, memory controller, and the memory cube network.
It periodically takes the new state of the system environment and generates the action accordingly to either remap data or computation.
If a data remapping action is taken, the virtual to physical mapping of a highly accessed page is updated in the paging system by the OS, meanwhile it is migrated from its current cube to a new cube in the memory.
If a compute remapping action is taken, future NMP operations related to a highly accessed page is directed to a new location.
To evaluate the impact of the mapping decisions, the performance monitor in the memory controller continuously observes the feedback from the memory and calculates the reward for the previous actions, which is used to train and improve the RL agent.

\begin{figure}
    \centering
    \includegraphics[width=\columnwidth]{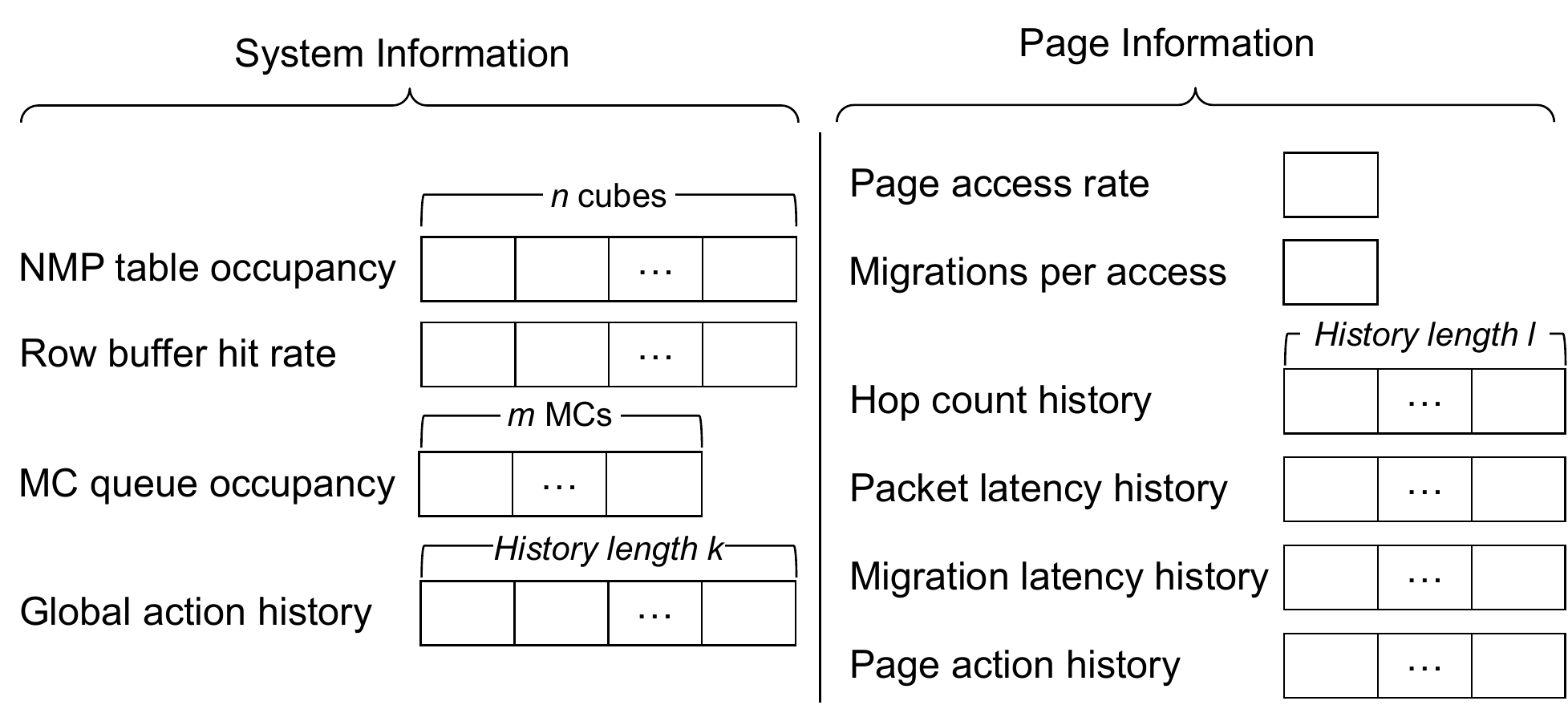}
    \caption{The state representation of \tech{} consists of system and page information.}
    \vspace{-0.5cm}
    \label{fig:state}
\end{figure}

\subsection{NMP Memory Mapping Representation}
\label{rl_solution_approach}

Proper representations for the state, action and reward are important but challenging to derive for reinforcement learning.
We describe the considerations in the context of NMP and detail the derived representations through our empirical study as follows.

\textbf{State Representation.} The environment of an NMP system consists of the hardware system and the running application.
Therefore, we capture both system and application information to represent the state.
Figure~\ref{fig:state} shows the set of information that represents a state.
For the system information, we collect the NMP computation resources such as \textit{operation buffer occupancy}, and \textit{average row buffer hit rate} for each memory cube.
In addition, each \textit{memory controller queue occupancy} is also recorded for monitoring the regional congestion of the memory network.
Moreover, a \textit{history of previous actions} is also included to provide more information. For tracking the application behavior, we select the highly accessed page as a representative of the workload and collect the page specific information including computation and memory access patterns.
The selected page is also a target for memory mapping, which remap either the page or its associated computation to a new cube. 
The \textit{migration latency history} and the \textit{action history} for the page are part of the page information.
In addition, the \textit{page access rate} with respect to all the memory accesses and the \textit{migrations per access} of the page are recorded.
Several communication metrics of the past related NMP operations are also monitored.
Moreover, a fixed length history of communication hop count and round-trip latency of the data in the page is selected to reflect the application behavior.

\textbf{Action Representation.} We have in total eight actions, including six for data or computation remapping and two for adjusting the agent invocation interval., 
The interval-related actions are meant to adjust the agent's learning interval based on an application's unique behavior.
The discrete intervals used in this work are 100, 125, 167, and 250 cycles.
The actions are summarized as follows.
\begin{enumerate}[topsep=0.3em,itemsep=-0.2em,label=(\roman*)]
    \item \textit{Default mapping}: no change in the mapping.
    \item \textit{Near data remapping}: remap the page to one of the compute cube's neighbor cubes randomly.
    \item \textit{Far data remapping}: remap the page to the compute cube's diagonal opposite cube in the 2D cube array.
    \item \textit{Near compute remapping}: remap the compute location to one of the current compute cube's neighbors.
    \item \textit{Far compute remapping}: remap the compute location to the current compute cube's diagonal opposite cube.
    \item \textit{Source compute remapping}: remap the compute location to the host cube of first source data of involved NMP operations.
    \item \textit{Increase interval}: increase agent invocation interval.
    \item \textit{Decrease interval}: decrease agent invocation interval.
\end{enumerate}

\textbf{Reward Function.} The reward function returns a unit of positive (+1) and negative (-1) reward for performance improvement or degradation.
Otherwise, a zero reward is returned.
We've explored using the communication hop count as a performance metric, but it leads to a local minimum that does not reflect the performance goal.
We empirically found that operations per cycle as a direct reflection of performance can achieve a robust learning process. Therefore, it is used as the performance metric in the reward function.


\subsection{RL Agent}

We use the off-policy, value-based deep Q-learning~\cite{mnih2015human} algorithm for the proposed RL-based \tech{}.
For the state-action value estimation, we use a dueling network as a function approximator.
The DNN model in the agent is a simple stack of fully connected layers, as shown in Figure~\ref{fig:sys_flow_arch}~\circled{3}.
The agent takes the state and predicts the state-action value for each action.
Then we use an $\epsilon$-greedy Q-learning algorithm~\cite{sutton2018reinforcement} to trade off the exploitation and exploration during the search and learning process.
The algorithm would select an action randomly with probability $\epsilon$ to explore the decision space, and choose the action with the highest value with probability $1-\epsilon$ to exploit the knowledge learned by the agent.
To train the DNN, we leverage \textit{experience replay}~\cite{mnih2015human} by keeping the past experiences in the replay buffer and randomly draw the samples for training.
Therefore, the learning and search process is more robust by consolidating the past experiences into the training process.

\section{Hardware Implementation}
\label{hardware_implementation}
\subsection{Information Orchestration}
\label{information_orchestration}


We propose hardware to orchestrate the system and page specific information for providing system and application state as well as rewards to the agent.
In each memory cube, we deploy an NMP table for keeping the outstanding NMP operations (Ops).
Additionally, registers are used for tracking the NMP table occupancy status and the average row buffer hit rate in each cube.
These two pieces of information are communicated to a cube's nearest memory controller (MC) periodically.
In each MC, two vectors of system information counters are used to record the NMP table occupancy and average row buffer hit rate of its nearest cubes, respectively.
Each counter saves the running average of the received value the MC receives from the corresponding cube.
A global history of actions is also tracked in the RL agent.

To maintain the page specific information for applications, a fully associative page information cache is added in each MC to collect the related information.
Each entry records the number of page accesses, number of migrations, and four histories, including the communication hop count, packet latency, migration latency, and actions taken for a page.
Upon sending NMP-op from MC to memory, accesses and hop count history of the entries of involving pages are updated.
If a page does not have a cache entry, the least frequently used policy is used to find one, which is cleared before updating.
Different from conventional cache, the content of the victim entry is abandoned.
When the MC receives an NMP acknowledgement (ACK), packet latency carried in the ACK is updated in the latency history of involving entries.
The migration latency history of an entry is updated after migration finishes if the corresponding page is remapped for migration.
Similarly, the action history is updated when the page is selected for the agent for an action.
The page access rate and migrations per access is calculated when forming a state for the agent to reduce the overhead.

Upon an agent invocation, the two system counter vectors from all the MCs with their MC queue occupancy, in addition to a global history of actions tracked in the RL agent are concatenated to form the system information.
Meanwhile, the page information of a highly access page is selected from one of the MC page information cache, where the MCs take turns to provide the page information in a round-robin fashion.
Then both system and page information are provided as a state to the agent for inference.
The performance value (operations per cycle) is also collected from all the MCs and compared with the previous one to determine the reward. 

\subsection{RL Agent Implementation}
\label{sec-rl-agent}
We propose to use an accelerator for deep Q-learning technique. Several such accelerators have been proposed in literature~\cite{fa3c, deepqaccelerator}. The block diagram of the RL agent is shown in Figure~\ref{fig:sys_flow_arch}. Periodically the agent pulls information from each memory controller. The incoming information includes the new state $s_t$ and a reward $r_{t-1}$ for the last state-action $(s_{t-1}, a_{t-1})$, are stored in the information buffer (\wcircled{1}).
The incoming information and the previous state and action in the information buffer form a sample $(s_{t-1}, a_{t-1}, r_{t-1}, s_t)$, which is stored in the replay buffer (\wcircled{2}).
Meanwhile, the agent infers an action $a_t$ on the input state $s_t$ (\wcircled{2}).
The generated action $a_t$ is stored back to the information buffer and transmitted back to the MCs which then perform the appropriate operations (\wcircled{3}).
Upon the training time, the agent draws a set of samples from the replay buffer for training and applies a back propagation algorithm to update the DNN model.

\subsection{Page and Computation Remapping}
\label{migration_implementation}

Actively accessed pages are chosen as candidates to incorporate into state information for the agent to evaluate their fitness in the current location under the current system and application state.
The agent may suggest page and computation remapping to adapt to the new application behavior and system state, which are implemented through page migration and NMP  Op scheduling, respectively.

 

\textbf{Page Remapping.} It involves OS for page table update and the memory network for page migration to reflect a page remapping, where the virtual page is mapped to a new physical frame belonging to a memory cube suggested by the agent.
We provide \textit{blocking} and \textit{non-blocking} modes for pages with \textit{read-write} and \textit{read-only} permissions, respectively.
For blocking migration, the page is locked during migration and no access is allowed in order to maintain coherence.
For non-blocking migration, the old page frame can be accessed during the migration to reduce performance overhead.
When a page migration is requested by the data remapping decision from the agent, the page number and the new host cube are put into the migration queue of the migration management system.
When the migration DMA can process a new request, the OS is consulted to provide a frame belonging to the new host cube.
Then DMA starts generating migration requests that request data from the old frame and transfer it to the new frame in the new host cube.
Once migration finishes, a migration acknowledgement is sent from the new host to the migration management system, which then reports the migration latency to the memory controller.
Meanwhile, an interrupt is issued to invoke the OS for page table update.
In case of blocking migration, the page is unlocked for accessing.
For non-blocking migration, the old frame is put back to the free frame pool when the outstanding accesses all finish while new accesses use the new mapping.

\textbf{Computation Remapping.} Computation remapping decouples computation location and the data location for balancing load and improving throughput of the NMP memory network.
Computation cube of an NMP-op is determined by the NMP-op scheduler based on the data address.
The computation cube is then embedded in the offloaded NMP-op request packet.
A compute remap table is used to remap the computation to a different cube suggested by the agent.
When a compute remapping decision related to a page is given by the agent, the page number and the suggestion are stored in the compute remap table.
Upon scheduling an NMP-op, the NMP-op scheduler consults the compute remap table.
If the related page of the NMP-op has an entry in the table, the computation cube is decided based on the agent suggestion recorded in the entry.
Otherwise, the default scheduling is used.


\section{Evaluation Methodology}
\label{sec:methodology}
We talk about the simulation methodology, system modeling, workload and workload characterization in this section.  

\subsection{Simulation Methodology}

We develop a framework by integrating an in-house cycle-accurate simulator with keras-rl~\cite{plappert2016kerasrl} and the \tech{} learning module built upon gym~\cite{brockman2016openai}.
The simulator is responsible for timing simulation while the gym-based \tech{} agent and keras-rl are used for functional simulation.
We used trace-driven simulation with NMP-Op traces collected from applications with medium input data size by annotating NMP-friendly regions of interest that were identified in previous works~\cite{ahn2016pei,huang2019active}.
The traces of an application form an episode for the application.
For single-program workloads, we run each application episode for 5 times, where each time simulation states are cleared except the DNN model.
For multi-program workloads, we run multiple (2, 3, and 4) applications concurrently for 10 times, where each new run clear the simulation states except the DNN model.
The combinations of multi-program workloads are decided based on the workload analysis (\S\ref{subsec:workload_analysis}).
For estimating the area and energy, we model all the buffers and caches using Cacti~\cite{balasubramonian2017cacti}, 45nm technology, since these components contribute to most of the area and energy overhead in the system.

\begin{table}[h!]
    \centering
    \resizebox{\columnwidth}{!}{
    \begin{tabular}{l|l}
        \hline\hline
        Hardware & Configurations  \\ \hline\hline
        Chip Multiprocessor (CMP) & {16 core, Cahce (32KB/each core), MSHR (16 entries)} \\ \hline
        Memory Controller (MC) & 4, one at each CMP corners, Page Info Cache (128 entries) \\ \hline
        Memory Management Unit (MMU) & {4-level page table} \\ \hline
        Migration Management System (MMS) & Migration Quaue (128 entries)\\ \hline
        Memory Cube & 1GB, 32 vaults, 8 banks/vault, Crossbar \\ \hline
        Memory Cube Network (MCN) &  4$\times$4 mesh, 3 stage router, 128 bit link bandwidth \\ \hline
        NMP-Op table & 512 entries \\ \hline
    \end{tabular}
    }
    \caption{Hardware Configurations.}
    \label{tab:architecture_parameter}
\vspace{-0.2cm}
\end{table}

\subsection{System Modeling}

We develop a simple programming interface for simulating memory allocation, light weight process creation and launching, MMU with 4-level page table, processor core and NMP operation loader, in addition to cache and MSHR.
The \tech{} modules highlighted in Figure~\ref{fig:sys_flow_arch} are also implemented and connected with memory controllers that are attached to four corner cubes of the memory network.
For the memory network, we model 4$\times$4 and 8$\times$8 mesh networks with 3D memory cubes that consists of vaults, banks and router switches.
We implement an NMP table for storing NMP-Ops and their operands, computation logic, and the router on the base die of the memory cube.
Each router has 6 ports and 5 virtual channels, to make sure that there is no protocol deadlock in the system, following static XY routing and token-based flow control for packet switching.
The architectural parameters are summarized in Table~\ref{tab:architecture_parameter}.

\begin{figure*}[t]
\centering
    \begin{subfigure}{0.34\textwidth}
    \includegraphics[scale=0.26]{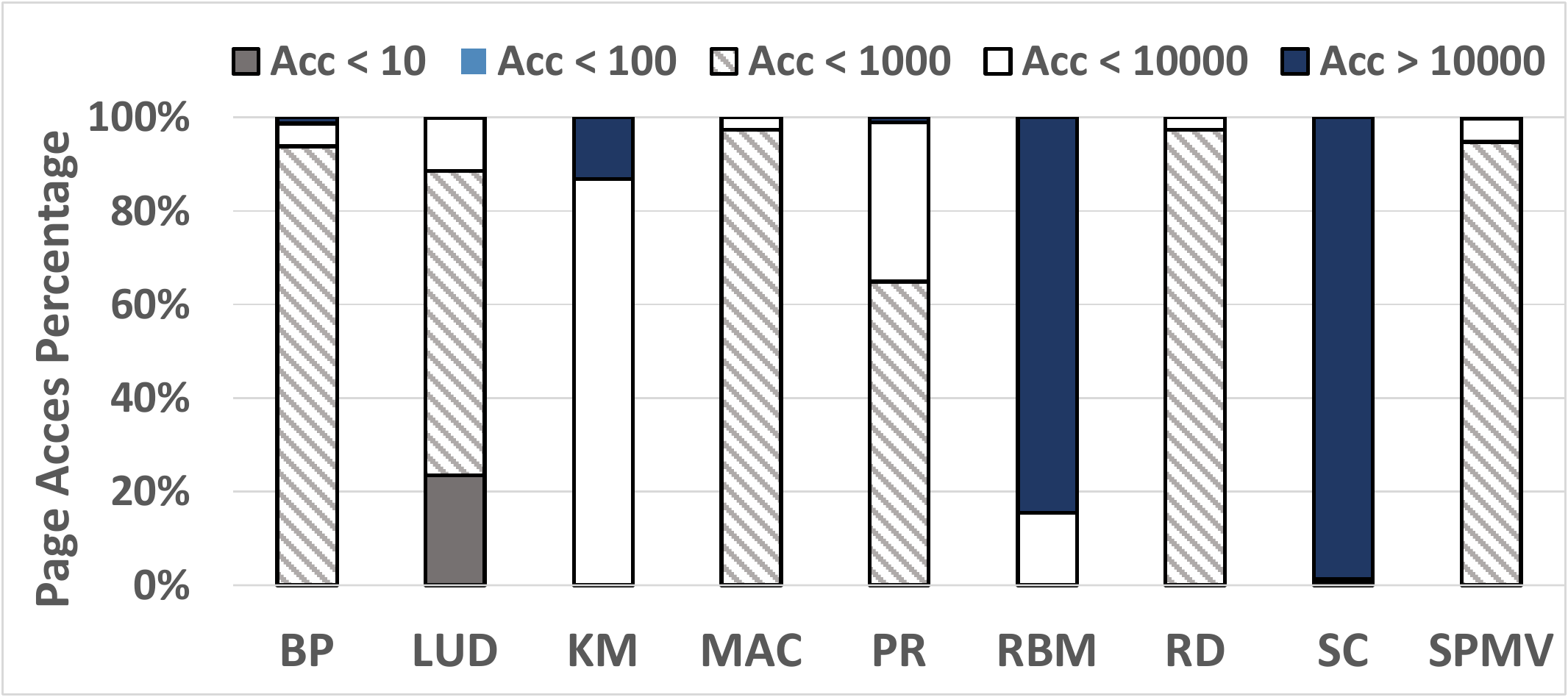}
    \caption{Classification of pages based on the number of accesses.}
    \label{fig:accesses_per_page}
    \end{subfigure}
    \begin{subfigure}{0.32\textwidth}
    \includegraphics[scale=0.38]{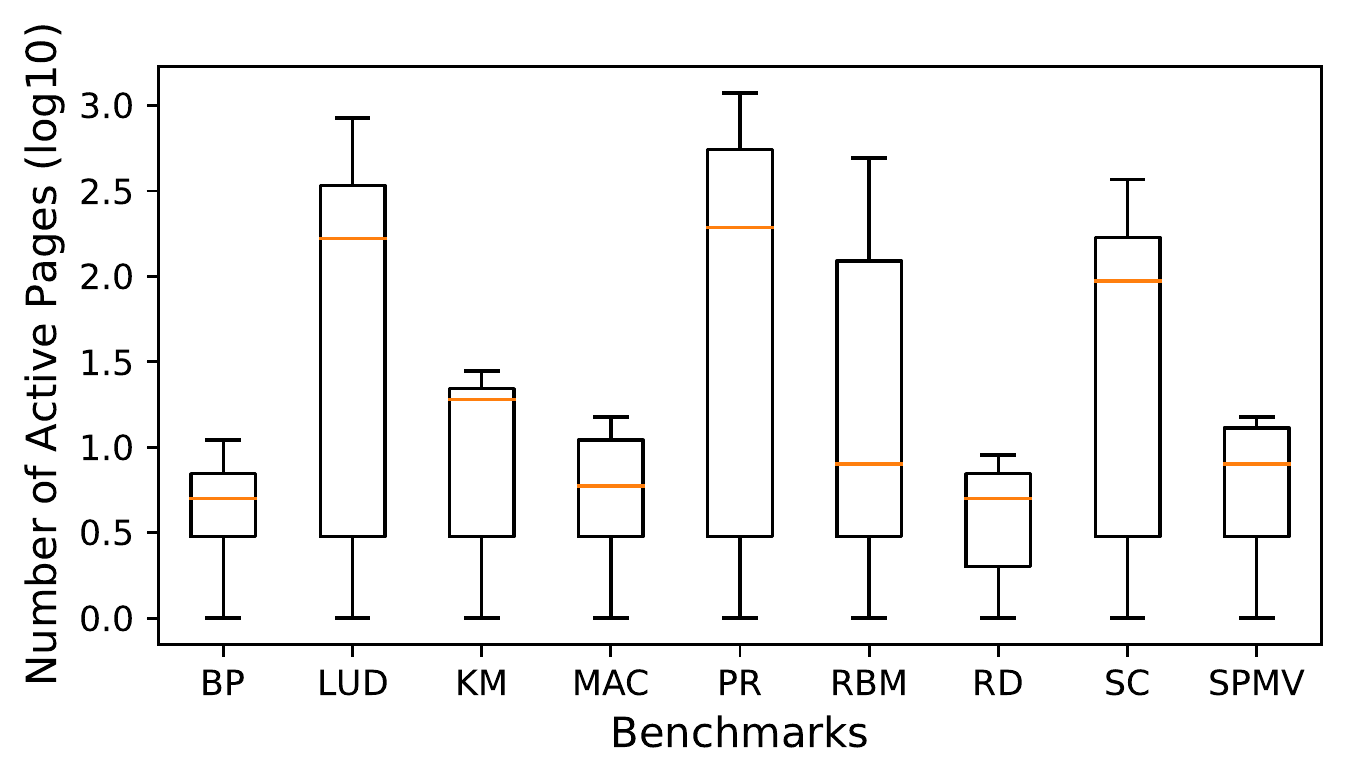}
    \caption{Active page distribution for each applications.}
    \label{fig:active_pages}
    \end{subfigure}
    \begin{subfigure}{0.33\textwidth}
    \includegraphics[scale=0.265]{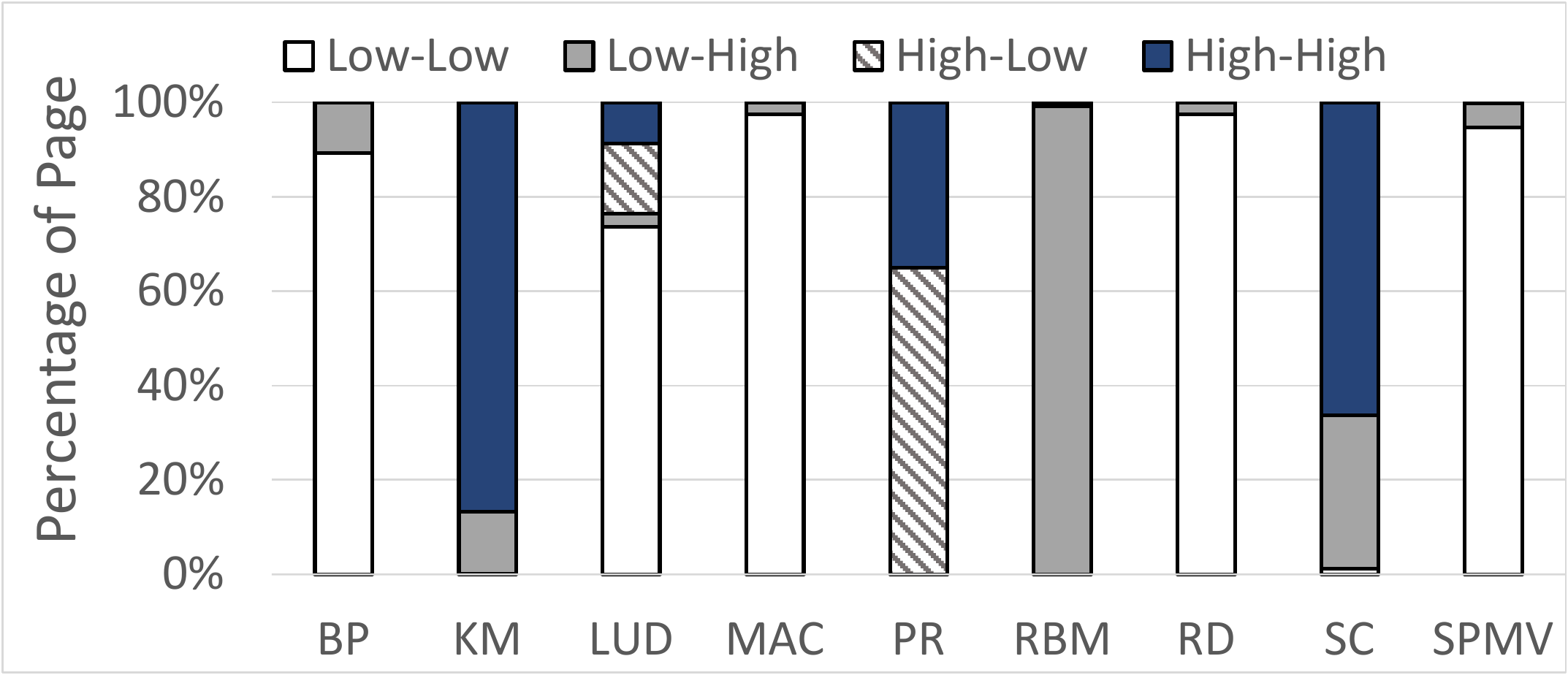}
    \caption{Page affinity analysis.}
    \label{fig:affinity_analysis}
    \end{subfigure}
    \vspace{-0.2cm}
    \caption{Workload Analysis Graphs. (a) Classification of pages based on their access volume, (b) Active Pages representing the working set, (c) Page Affinity showing the interrelation among the pages in an application.}
    \vspace{-0.2cm}
    \label{fig:workload_analysis}
\end{figure*}

\subsection{NMP Techniques and Mapping Schemes}
We have implemented (1) three NMP techniques, Basic NMP (BNMP), Load Balancing NMP (LDB), PIM Enabled Instruction (PEI)\cite{ahn2016pei}, (2) one physical address remapping technique, Transparent Offloading and Mapping (TOM)~\cite{hsieh2016tom}, and (3) state-of-the-art page-frame allocation policy HOARD~\cite{hoard}. We briefly discuss about each of them as follows. 

\textbf{Basic NMP (BNMP):} We implement BNMP, following  Active-Routing~\cite{huang2019active} for scheduling each operation in the memory cube while the in-network computing capability is not included. The NMP operation format is considered as \texttt{<\&dest += \&src1 OP \&src2>}, where \texttt{\&dest} page host cube is considered as the computation point. An entry is made in the NMP-Op table at the computation cube and requests are sent to other memory cubes if sources do not belong to the same cube as the destination operand page. Upon receiving responses for the sources, the computation takes place and the NMP-Op table entry is removed once the result is written to the memory read-write queue. The response is also sent back to the CPU if required.     

\textbf{Load Balancing NMP (LDB):} This is a simple extension of BNMP, based on two observations as follows. (1) Oftentimes, some NMP-Op table receives a disproportionate load based of the applications access pattern. (2) In most of the applications, the number of pages used for sources is significantly higher than the number of pages used for destination operands. Hence we simply change computation points from destination to sources in order to balance the load on the NMP-Op table. However, once the computation is done, the partially computed result must be sent back to the destination memory cube and also to the CPU. 

\textbf{PIM Enabled Instruction (PEI):} This technique recognizes and tries to simultaneously exploit the benefit of cache memory as well as NMP. In case of a hit in the cache for at least one operand, PEI offloads operation with one source data to another source location in the main memory for computation. 

\textbf{Transparent Offloading and Mapping (TOM):} This is a physical-to-DRAM address remapping technique, originally used for GPUs to co-locate the required data in the same memory cube for NMP.
Before kernel offloading, TOM profiles a small fraction of the data and derives a mapping with best data co-location, which is used as the mapping scheme for that kernel.
We imbibe the mapping aspect of TOM and make required adjustments to incorporate it in our context for remapping data in the NMP system.
We infer data co-location from data being accessed by NMP-Op traces. Each mapping candidate is evaluated for a thousand cycles with their data co-location information recorded.
Then the scheme with best data co-location that incurs the least data movement is used for an epoch.

\textbf{HOARD:}
We adopted an NMP-aware HOARD~\cite{hoard} allocator as the baseline, heuristic-based OS solution for comparison and also as a foundation for experiment.
HOARD is a classic multithreaded page-frame allocator which has inspired many allocator implementations~\cite{ptmalloc, tcmalloc, jemalloc}
in modern OSes.
The original version of HOARD focuses on improving the temporal and spatial locality within a multi-threaded application.
HOARD maintains a global free list of larger memory chunks which are then allocated for each thread to serve the memory requests of smaller page frames or objects.
Once the thread has finished the usage of the memory space, it chooses to ``hoard'' the space in a thread-private free list until the space is reused by the same thread or the whole chunk gets freed to the global free list.
As a result, HOARD is able to co-locate data that belongs to same thread as much as possible.
We adapted the thread-based heuristic of HOARD for each program in our multi-program workload setting.
Our HOARD allocator aims for improving the locality within each program,
contributing to the physical proximity of data that is expected to be accessed together in the NMP system.


\subsection{Workloads}
We primarily target the long running applications with large memory residency, which repeatedly use their kernels to process and compute on huge numbers of inputs. The machine learning kernels are a natural fit for that as they are widely used to process humongous amounts of data flowing in social media websites, search engines, autonomous driving, online shopping outlets, etc. These kernels are also used in a wide variety of applications such as graph analytics and scientific applications. To mimic realistic scenarios, we create both single and multi-program workloads using these application kernels. Since these are well known kernels, we describe them briefly in Table~\ref{tab:benchmarks}.     

\begin{table}[h!]
    \centering
    \resizebox{\columnwidth}{!}{
    \begin{tabular}{l|l}
        \hline\hline
        Benchmarks & Description  \\ \hline\hline
        Backprop (BP)~\cite{che2010characterization} & Widely used algorithm for training feedforward neural networks. \\
         & It efficiently computes the gradient of the loss function with 
         respect \\ &to the weights of the network.
                               \\ \hline
        Lower Upper  & In numerical analysis factors a matrix as the  
        product of a lower \\ decomposition (LUD)~\cite{che2010characterization} & triangular matrix and an upper triangular matrix. \\ \hline
        Kmeans (KM)~\cite{che2009rodinia} & Kmeans algorithm is an iterative algorithm that
        tries to partition the \\ 
        &data-set into K pre-defined distinct non-overlapping 
        subgroups (clusters) \\
        &where each data point belongs to only one group. 
                                      \\ \hline
        \multirow{1}{*}{MAC}& multipy-and-accumulate over two sequential vectors.\\ \hline
        \multirow{2}{*}{Pagerank (PR)~\cite{ahmad2015crono}}& PageRank works by counting the number and quality of links to a page\\ & to determine a rough estimate of how important the website is.   \\ \hline
                            
        Restricted Boltzmann  & RBMs are a variant of Boltzmann machines\cite{hinton2007boltzmann}, with the restriction that \\ Machine (RBM)\cite{thomas2014cortexsuite}  &their neurons must form a bipartite graph. \\ \hline
        \multirow{1}{*}{Reduce (RD)} & sum reduction over a sequential vector. \\\hline
        \multirow{2}{*}{Streamcluster (SC)~\cite{bienia2008parsec}} & It assigns each point of a stream to its nearest center Medium-sized \\ &working sets of user-determined size. \\ \hline
        Sparse matrix-vector &  It finds applications in iterative methods to solve sparse linear systems and \\ multiply (SPMV)~\cite{che2009rodinia} &information retrieval, among other places. \\ \hline
    \end{tabular}}
    \caption{List of benchmarks.}
    \label{tab:benchmarks}
\vspace{-0.2cm}
\end{table}

\subsection{Workload Analysis}
\label{subsec:workload_analysis}
To capture the page-frame mapping related traits of the application, we characterize the workloads in terms of (1) page usage as an indicator of page life-time, (2) number of pages actively used in an epoch as an indicator of working set, and (3) inter-relation among the pages (page affinity) accessed to compute NMP operations in order to analyze the difficulty level for optimizing their access latency. 
\begin{figure*}[t]
    \centering
    \includegraphics[scale=0.45]{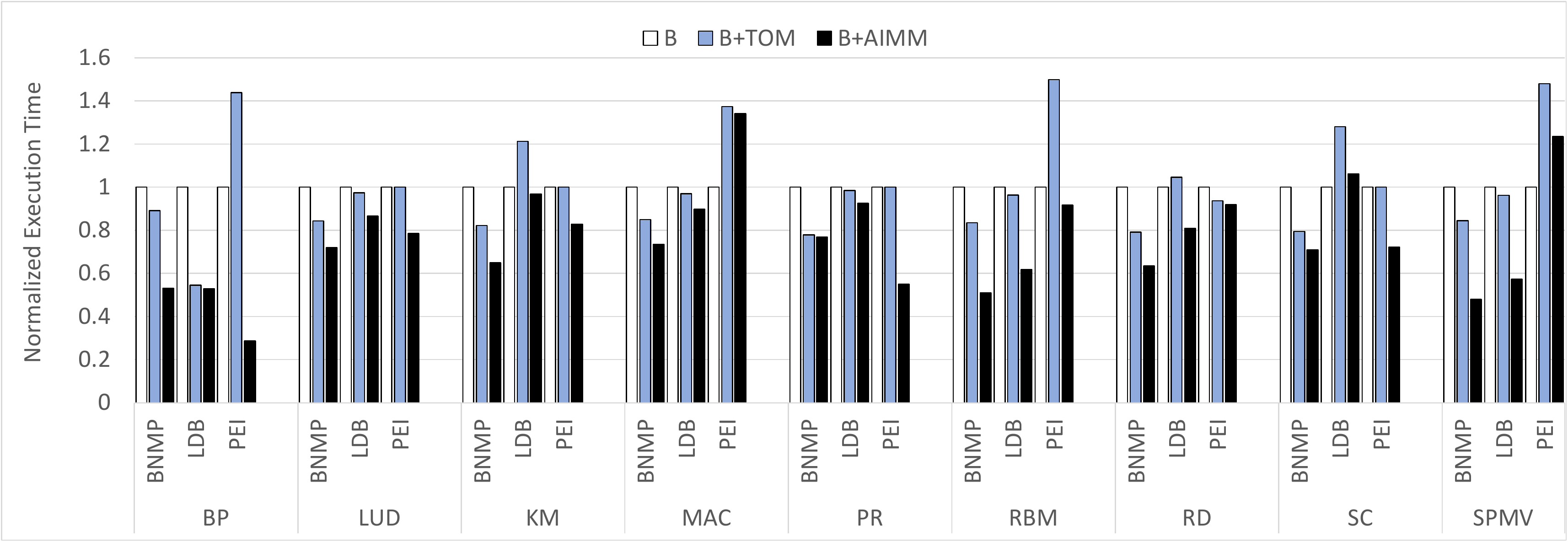}
    \caption{Execution time for all the benchmarks, normalized individually with their basic techniques BNMP, LDB, and PEI respectively, which does not have any remapping support, and commonly referred as B in the graph. Execution time for the baseline implementations are compared to the system with remapping support, namely, TOM and \tech, respectively.}
    \label{fig:execution_time_1p}
\end{figure*}

\textbf{Page Access Classification:}
The page usage is an indicator of its scope for learning the access pattern and improving performance after data or computation remapping at runtime. If the number of accesses to pages are very low, the scope for improvement using remapping technique narrows down to a great extent. In Figure~\ref{fig:accesses_per_page}, we show that for most of our application kernels, the majority of the pages are moderate to heavily used, which offers a substantial scope for \tech. 

\textbf{Active page distribution:} It is defined as the pages that are being accessed in a fixed time window (epoch). We collect it for each epoch separately and do an average for producing the active page numbers, as shown in Figure~\ref{fig:active_pages}. We can clearly identify two classes of applications. (1) High number of active pages, \textit{LUD, PR, RBM, SC}, and (2) Low or moderate number of active pages, \textit{BP, KM, MAC, RD, SPMV}. This study gives us an indication of the amount of page information, ideally needed to be stored in the page information cache for the training of the agent.   


\textbf{Affinity analysis:} In data mining, affinity analysis uncovers the meaningful correlations between different entities according to their co-occurrence in a data set. In our context, we use and define the affinity analysis to reveal the relationship among different pages if they are being accessed for computing the same NMP operation. We track two distinct yet interlinked qualities of page access pattern, (1) the number of pages related with a particular page as the radix for that page, which is similar to radix of the node in a graph, (2) the number of times each of those pair of connected nodes are accessed as part of the same NMP operation, similar to the weight for each edge. To understand the affinity, we create N number of bins for each of the traits and place the pages in the intersection of both by considering the traits together. So the affinity space is $N (accesses)\times N (radix)$, which is further divided into four quadrants to produce a consolidated result in Figure~\ref{fig:affinity_analysis}. Based on our study, a higher affinity indicates a harder problem, which poses greater challenges for finding a near-optimal solution. On the contrary, they also exhibit and offer a higher degree of scope for improvement. Please note that this is only one aspect of the complexities of the problem. Interestingly in our collection, we observe a balanced distribution of workloads in terms of their page affinity.

\section{Experimentation Results}
\label{results}

\begin{figure}[t]
    \centering
    \includegraphics[scale=0.31]{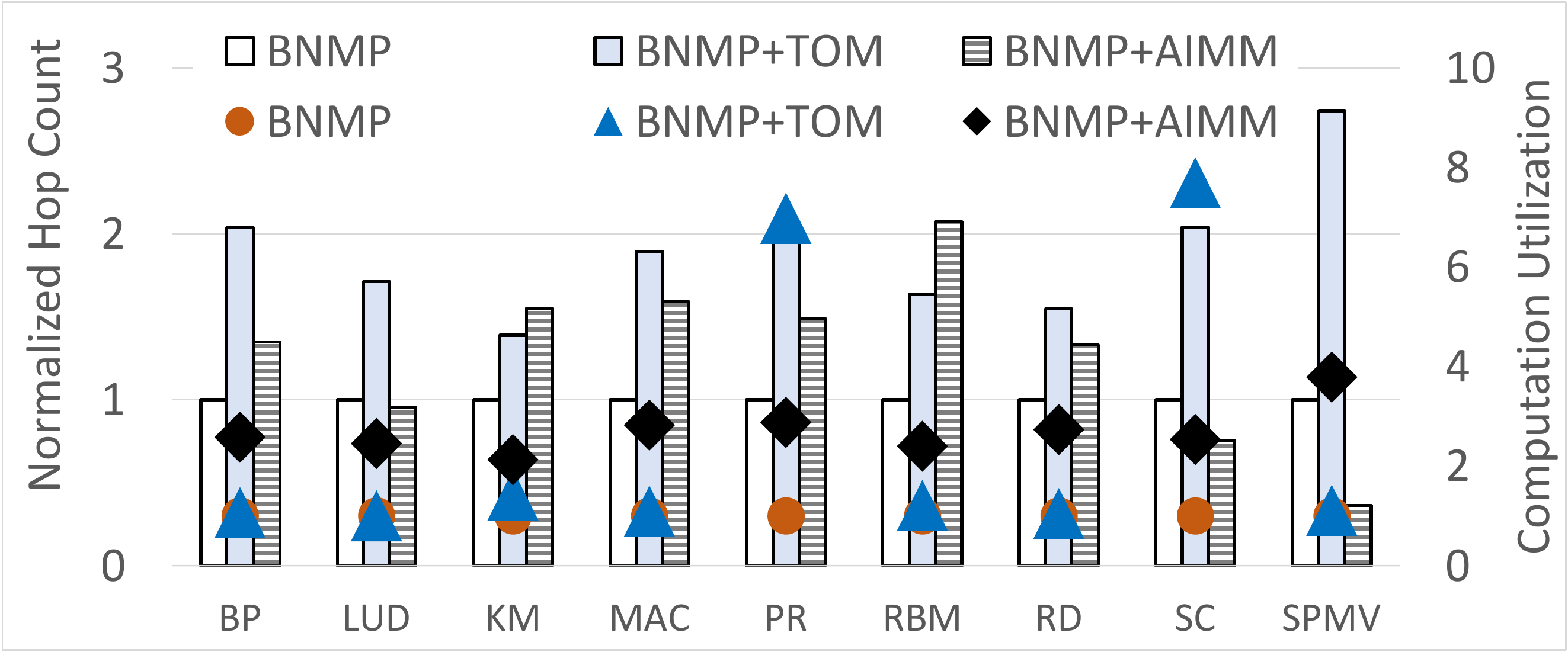}
    \caption{Average Hop Count and Computation Utilization. Major Y-axis is shown in bars.}
    \label{fig:hop_count_compt_util}
\end{figure}

\subsection{Performance}
\subsubsection{Execution Time}
In Figure~\ref{fig:execution_time_1p},
we show the execution time for different applications under various system setup and support. 
It is evident that \tech~is effective at helping the NMP techniques to achieve better performance in almost all cases. We observe up to 70\% improvement in execution time. With BNMP processing, the NMP operations are completely on the memory side, both TOM and \tech~boost the performance of BNMP. In general, TOM achieves around 15\% to 20\% performance improvement.
\tech~ secures improvement in execution time by around 50\% on average across all the benchmarks. However for LDB, TOM did not improve much for most of the benchmarks, with \textit{BP} as an exception. With LDB, \tech~improves execution time up to 43\%, with no performance degradation observed except minor performance loss for \textit{SC}. In the case of PEI, it is evident that for the majority of benchmarks TOM degrades the performance, whereas \tech~manages to achieve performance benefit around 10\% to 20\% (up to 42\%) on average. For \textit{SPMV} and \textit{MAC}, both TOM and \tech~ degrade the performance of PEI. 

There are several factors that play major roles in our system to drive system performance, such as operation per cycle, learning rate, utilization of migrated pages, path congestion and computation level parallelism. In addition, hop count in the memory network and row buffer hit rate in the memory cube can be considered as primary contributors for performance, depending on the nuances of individual cases. In the following, we discuss each of the techniques to justify their performance. 

In the case of \textit{PR} on BNMP, \tech~does not improve over TOM, which can mostly be attributed to the 3$\times$ computation distribution achieved by TOM over \tech~as shown on  Figure~\ref{fig:hop_count_compt_util}, in addition to very low migration page access for \textit{PR} as shown in Figure~\ref{fig:migr_stats}. Figure~\ref{fig:accesses_per_page} shows that \textit{PR} has high number of pages that accessed small number of times, justifying low usage of migrated pages. Higher computation distribution improves NMP parallelism at the cost of extra communication if the computation node and operation destination are different. Low accesses to migrated pages diminishes the benefit of migration. On the other hand, \tech~achieves 50\% better performance for \textit{SPMV} that is ascribed to significant improvement in average hop count in Figure~\ref{fig:hop_count_compt_util}, moderate fraction pages migrated (40\%) and they constitute almost 60\% of all the accesses as shown in Figure~\ref{fig:migr_stats}. 
The performance of \textit{SC} with TOM and \tech~allows us to delve into discussion of the trade-off between computation distribution and hop count in the memory network, along with importance of distribution of the right candidate in the right place. Missing one of them may offset the benefit achieved by the other, as the case for \textit{SC} with TOM. On the contrary, \tech~leads to a better performance than TOM with very moderate compute distribution and low hop count as shown in Figure~\ref{fig:hop_count_compt_util}.   

Since LDB and PEI both tend to co-locate the sources belonging to an NMP operation, they leave much lesser chance than BNMP type techniques for further performance improvement using remapping like TOM or \tech. The performance improvement is mainly achieved by optimizing the location of the destination pages with respect to the source operand pages. 


\begin{figure}[t]
    \centering
    \includegraphics[scale=0.26]{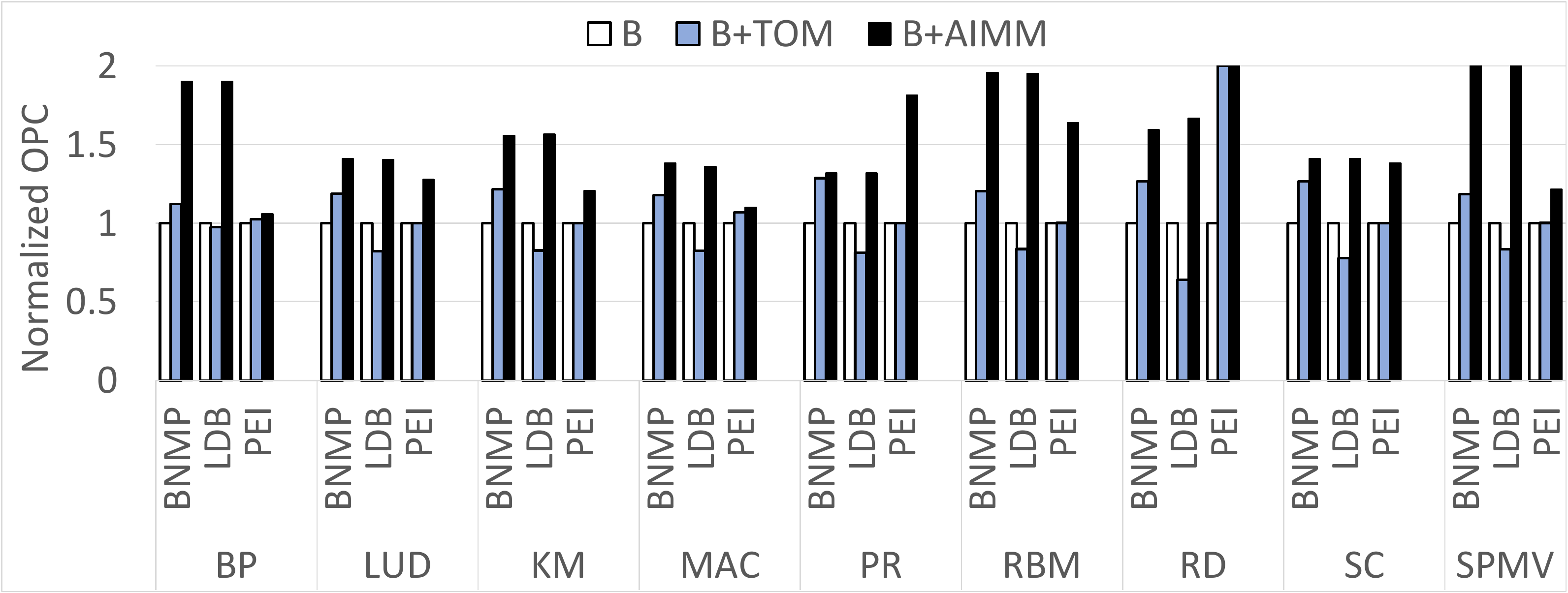}
    \caption{Normalized memory operations per cycle (OPC).} 
    \label{fig:norm_opc}
\end{figure}

\begin{figure}[t]
    \centering
    \includegraphics[scale=0.5]{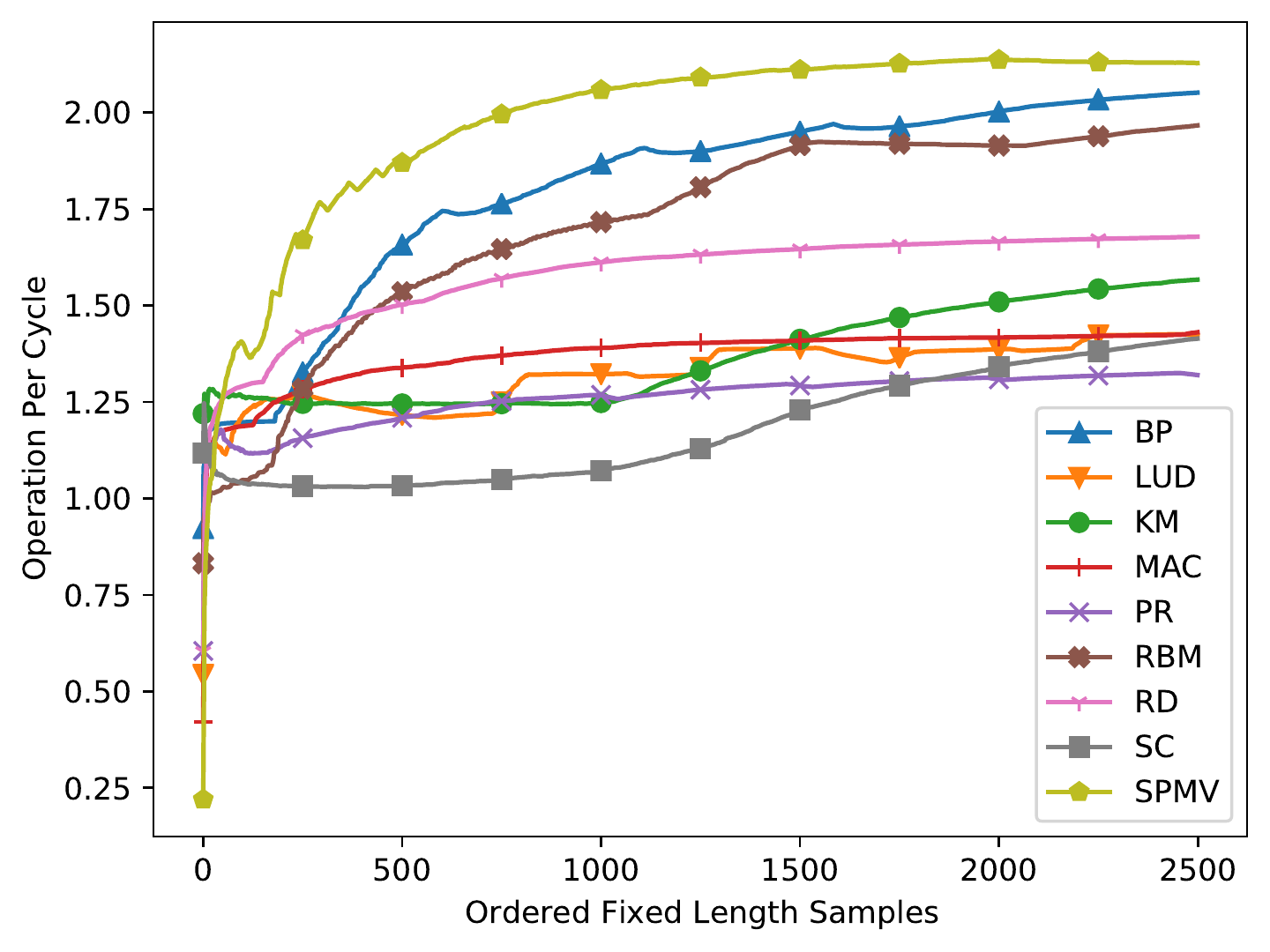}
    \caption{Operation per cycle timeline.}
    \label{fig:opc_timeline}
\end{figure}

\subsubsection{Operation Per Cycle (OPC)}
Similar to Instruction Per Cycle (IPC), in our experimentation for memory operations we coined the term Operation Per Cycle (OPC) as an indicator of the system throughput as shown in Figure~\ref{fig:norm_opc}. Since the goal of the \tech{} agent is to improve the OPC of the system, we observe significant improvement in OPC for all the benchmarks across all the techniques, except \textit{BP} and \textit{MAC} with PEI. This is because majority of pages are low-affine in those benchmarks, which can lead to less efficient learning outcome while carrying out actions that cause overhead. Since in the case of PEI+AIMM we also include the migration accesses while calculating OPC, if the migrations are not helpful, OPC improvement as shown in Figure~\ref{fig:norm_opc} may not directly translate to improvement in execution time. For instance, \textit{SPMV} has improved OPC for PEI, but suffered performance loss in terms of execution time, which suggested that higher OPC is caused by frequent but useless migrations in that particular case. Another interesting case is \textit{RD} with PEI, where OPC improved by almost 2$\times$, leading to only 5\% performance improvement both with TOM and \tech. By analyzing the results in detail, we observe that when \textit{RD} executes with both of these techniques, the row buffer miss rate increases significantly resulting in 42.68\% increase in DRAM access latency. This in turn slows down the network as well as the command scheduling from MC. In addition, we also observe more than 30\% inflation in average number of hops in both of these techniques, which contribute to higher latency.

\subsection{Learning Convergence}
In Figure~\ref{fig:opc_timeline}, we plot the timeline for OPC to show that \tech~progresses towards the goal of achieving high OPC as the time advances for each of the applications.\footnote{Since the number of operations are different for each of the applications, number of samples collected for each of the applications are also different. To plot them in the same graph we sampled them to a fixed size while preserving their original order.} The convergence signifies that the agent can learn from the system and is able to maintain steady performance once converged. As the system situation keeps changing, we constantly reevaluate the system situation and dynamically remap the page and computation location in the memory network in order to achieve near-optimum performance at any point of time.   

\begin{figure}[t]
    \centering
    \includegraphics[scale=0.32]{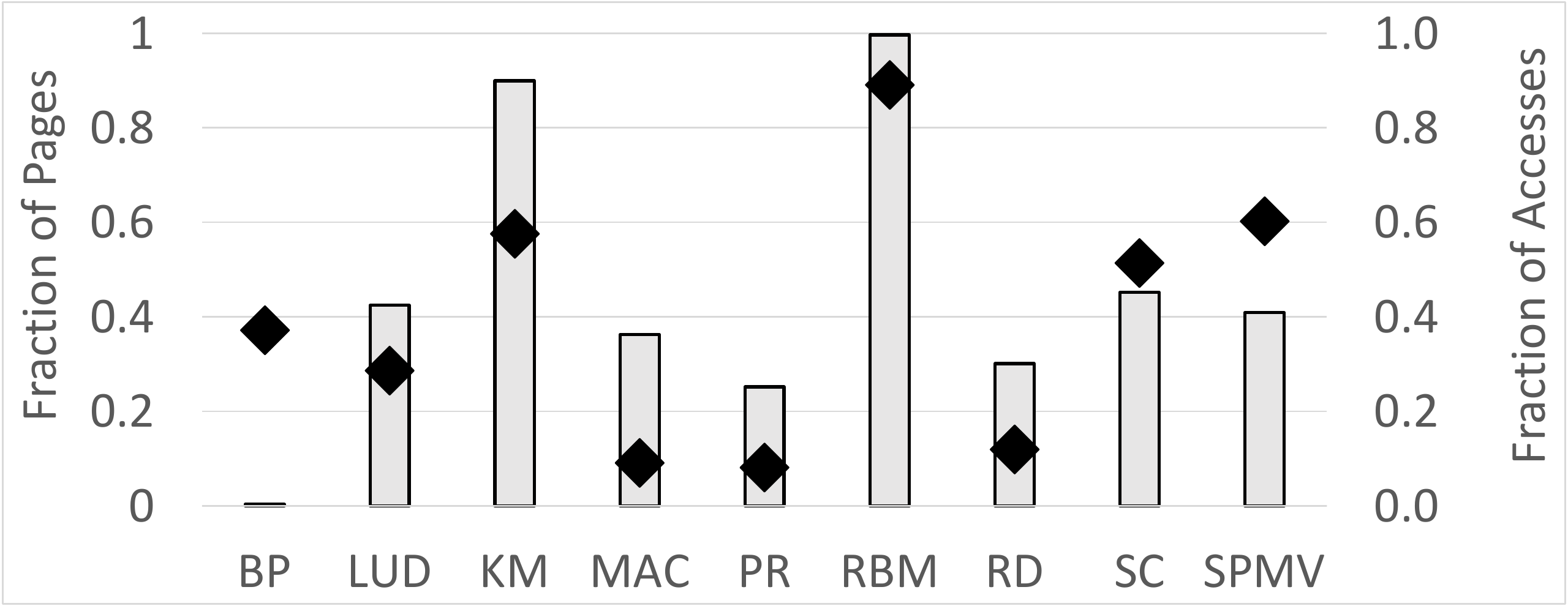}
    \caption{Migration Stats: On the major axis we show the fraction of pages that are migrated for each of the applications using bars. On the minor axis it projects the fraction of total accesses that are happened on migrated pages, with diamond shaped markers.}
    \label{fig:migr_stats}
\end{figure}

\subsection{Migration}
In Figure~\ref{fig:migr_stats}, we depict the fraction of pages being migrated for each applications on the major Y-axis and the fraction of total accesses requested that belong to a migrated page on the minor Y-axis. The fraction of pages migrated is an indicator of the migration coverage. For instance, in the case of \textit{RBM} 100\% pages are migrated, and almost all the migrated pages get accessed later. 
Pages in \textit{RBM} are susceptible to experience high volume of migration as (1) small number of pages are accessed most of the execution time, (2) in a small fixed time window almost all the pages are being accessed.
On the other hand, \textit{BP} has huge memory residency and relatively small working set, which leads to a low fraction of page migration as compared to the total number of pages. Interestingly, the small number of migrated pages constitute almost 40\% of the total accesses, which is a near ideal scenario as low number of page migrations has a small negative impact on performance, and a high number of accesses to the migrated pages can potentially improve the performance, provided the decision accuracy is high.  

\subsection{Hop Count and Computation Utilization}
In terms of system performance, hop count and computation distribution hold a reciprocal relation as reducing hop count improves the communication time, but results in computation under-utilization due to load imbalance across cubes. On the other hand, computation distribution may result in a high degree of hop count as concerned pages can only be in their respective memory cubes. Hence, a good balance between these factors is the key to achieving a near-optimal solution. Figure~\ref{fig:hop_count_compt_util} shows that \tech~ maintains a balance between the hop count and computation utilization. For instance, \textit{PR} and \textit{SC} achieves very high computation utilization which also leads to high average hop count. Assuming that the computation distribution decisions and corresponding locations are correct, the benefit gets diminished to 22\% and 20\% respectively possibly because of high hop count.  
 
\begin{figure}[t]
    \centering
    \includegraphics[scale=0.34]{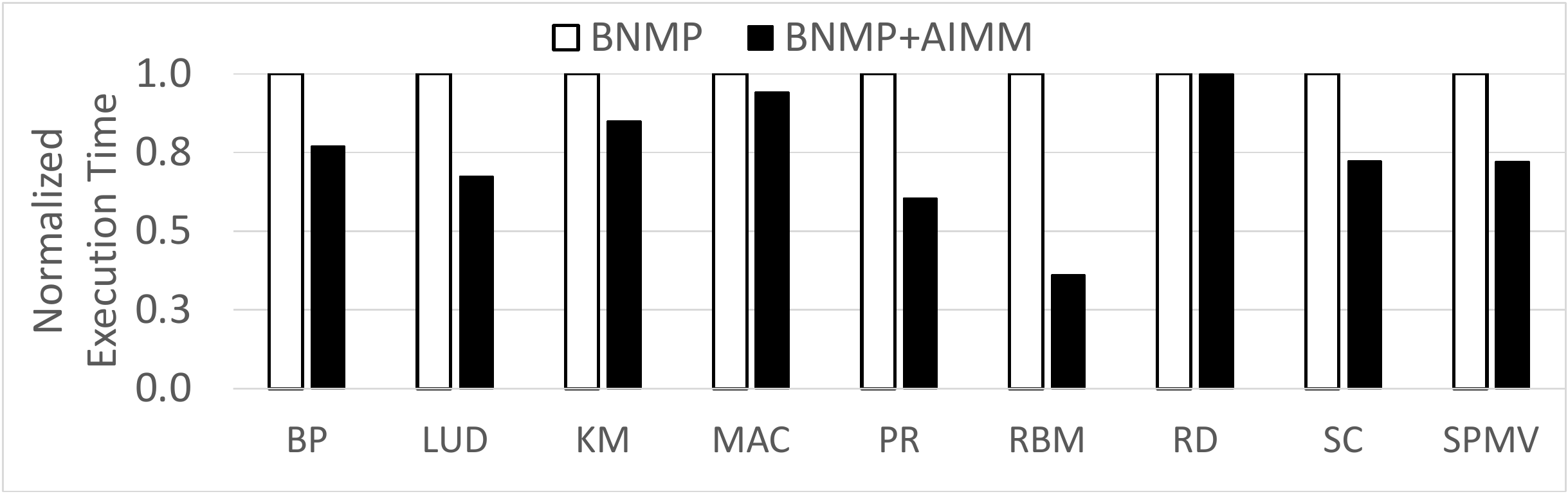}
    \caption{Normalized execution time for 8x8 mesh.}
    \label{fig:8x8_mesh}
\end{figure}

\subsection{Scalability Study}
In this subsection, we extend our experiments to study the impact of \tech~under a different underlying hardware configuration (8$\times$8 mesh) as well as under highly diverse workloads together (2/3/4-processes). With a larger network, we expect to observe higher network impact on the memory access latency, and intend to show that \tech~can adapt to the changes in the system without any prior training on them. While choosing applications for multi-program workloads, we select diverse applications together based on our workload analysis, so that the \rl~agent experiences significant variations while trying to train and infer with them.  

\subsubsection{MCN Scaling}

We observe that \tech~can sustain the changes in the underlying hardware by continuously evaluating them and without having any prior information. However, the amount of improvements for the applications are different than that in a 4$\times$4 mesh. For instance, with BNMP+\tech, \textit{RBM} observes more benefit over BNMP, than it observes in the 4$\times$4 mesh. As we know larger networks are susceptible to higher network latency, however, they also offer higher capacity and potential throughput. In the case of \textit{RBM}, \tech~ could sustain throughput, whereas benefits for other benchmarks slightly offset, mostly because of higher network delay.
Note that in terms of simulation with larger network size, we did not change the workload size. Tuning hyper parameters is left for future work.  

\begin{figure}[t]
    \centering
    \includegraphics[scale=0.285]{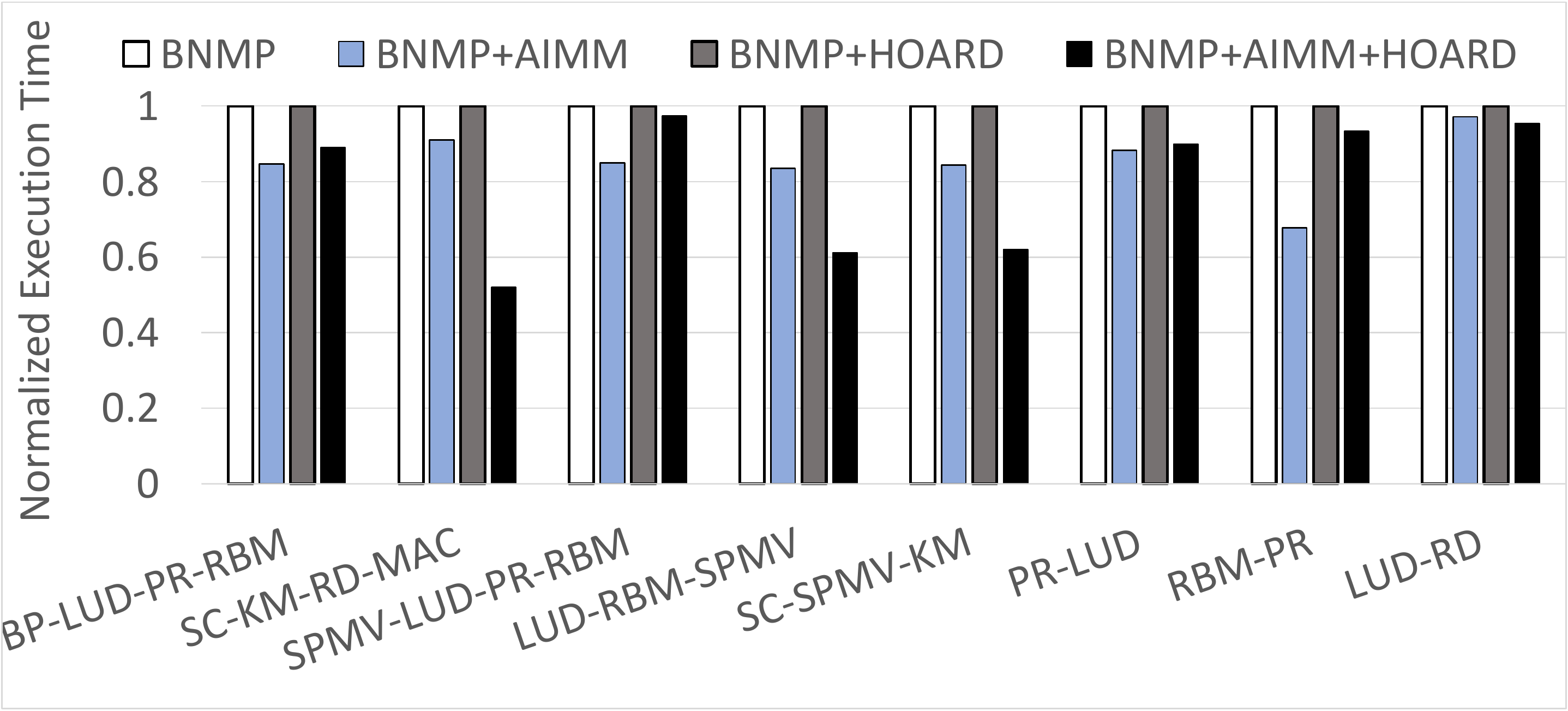}
    \caption{Multi-process normalized execution time.}
    \label{fig:execution_time_mp}
\end{figure}

\subsubsection{Multi-program Workload}
For a continuous learning environment like \tech, multi-program workloads pose a tremendous challenge on the learning agent as well as on the system. For studying the impact of multi-program workloads, we consider two baselines. (1) BNMP, where the NMP operation tables, page info cache, etc., are shared and contended among all the applications. We leave the study of priority based allocation or partitioning for future study. (2) BNMP+HOARD, where at the page frame allocator level our modified version of HOARD helps to co-locate data for each process, preventing data interleaving across processes. We observe that for several application combinations (\textit{SC-KM-RD-MAC}, \textit{LUD-RBM-SPMV}, \textit{SC-SPMV-KM}), HOARD and \tech~compliment each other to achieve 55\% to 60\% performance benefits. 

\begin{figure}[t]
    \centering
    \includegraphics[scale=0.32]{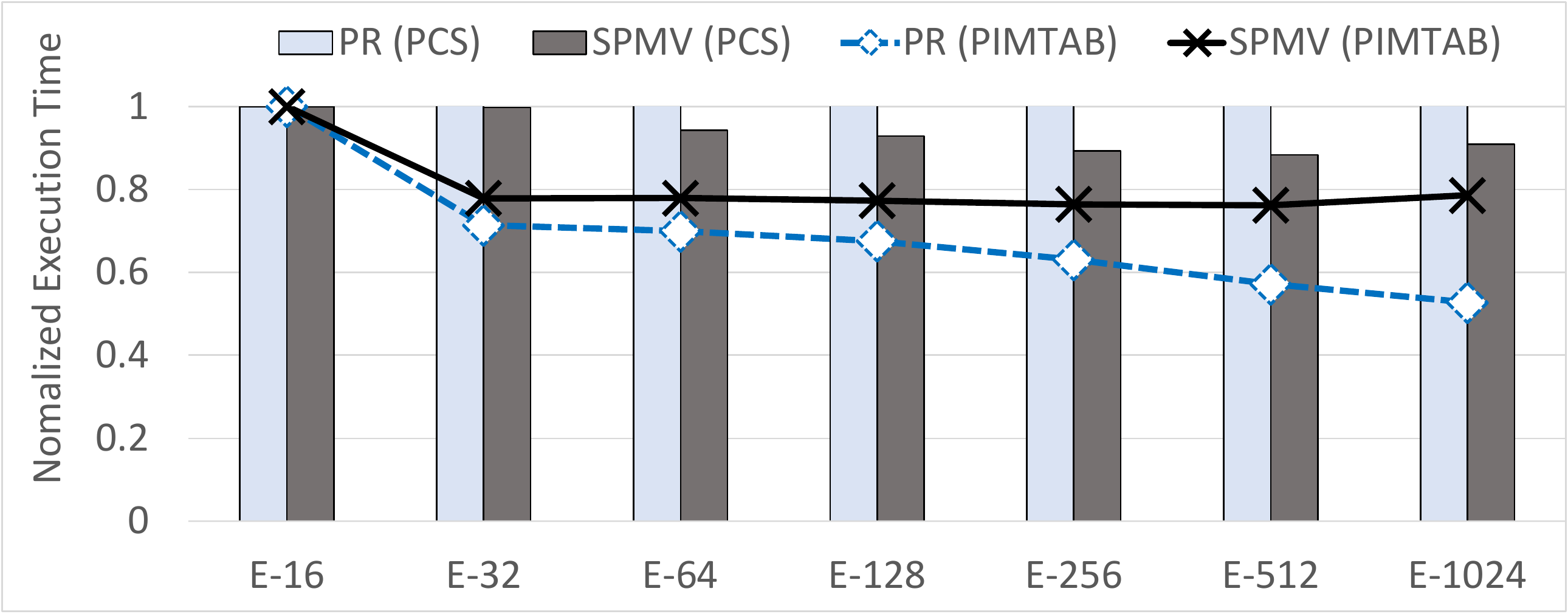}
    \caption{Sensitivity study: The bar graph shows the sensitivity of the benchmarks for different page-cache sizes, whereas the line graph show the sensitivity to the NMP-Op table sizes.}
    \label{fig:sensitivity}
\end{figure}

\subsection{Sensitivity Study}
We study the system performance by varying the sizes of (1) page info cache, whose size is critical for passing system information to the agent and (2) NMP operation table, whose size is important to hold the entries for NMP operations, denial of which affects memory network flow. We choose two representative applications (\textit{PR}, \textit{SPMV}) to study their performance by varying one parameter while keeping the other one to its default value, mentioned in Table~\ref{tab:architecture_parameter}.   

\textbf{Page Info Cache Size:} Figure~\ref{fig:sensitivity} shows that \textit{PR} exhibits very minimal sensitivity to page cache size, whereas \textit{SPMV} finds its sweet point while increasing the number of entry from 32 (E-32) to 64 (E-64). As a general trend, applications that get the most benefit from \tech~are more sensitive to the page cache size than others. Based on this study, we empirically decide the number of page cache entry as 256.

\textbf{NMP table size:} NMP table size sensitivity depends on several parameters such as the number of active pages (Figure~\ref{fig:active_pages}), computation distribution. In terms of active pages, \textit{SPMV} has around 10 active pages on average in a time window. Figure~\ref{fig:hop_count_compt_util} shows that \textit{SPMV} has the highest computation distribution among the other applications. Combining these two pieces of data, it explains the reason for execution time saturation after 32 entries (E-32) for \textit{SPMV}. On the other hand, \textit{PR} has a very high demand for NMP operation table, as it has high active page count and low reuse rate. That is why \textit{PR} shows monotonic improvement in execution time with the increase in the number of NMP table entries.  

\begin{figure}[t]
    \centering
    \includegraphics[scale=0.25]{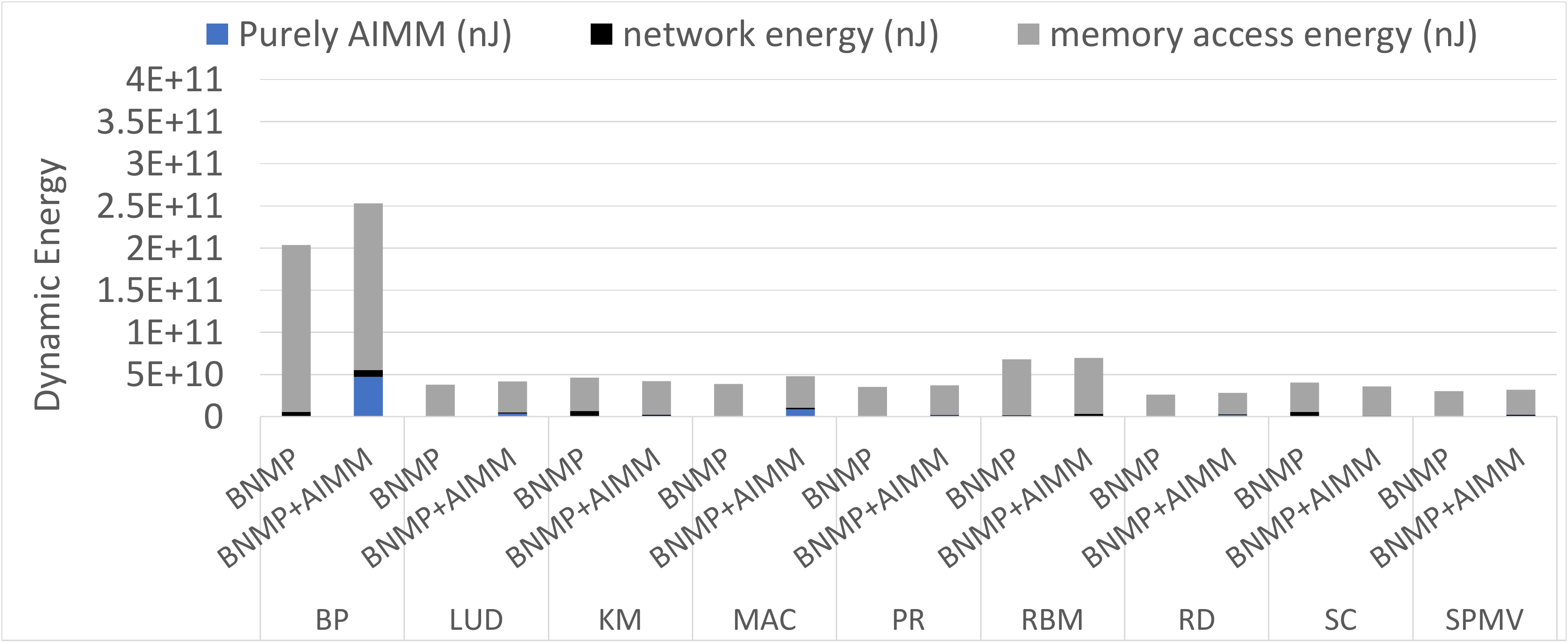}
    \caption{Dynamic energy (nJ) consumption.}
    \label{fig:energy}
\end{figure}

\subsection{Area and Energy}
In this section, we discuss the detailed area and energy aspect of our design. The implementation of the \rl~engine and migration management unit demands a separate provision in the chip. Additional energy consumption of information storing and propagation also contributes to the total energy consumption. We focus our study in five major design modules added for \tech{} and discuss the area and energy aspects for each of them separately. 

\textbf{(1) Information Orchestration:} We estimated the area for hardware registers and page information cache as part of the information orchestration system. Since the hardware registers occupy negligible area, we mostly focus on the page information cache of size 64KB, which occupies 0.23 mm$^2$ area. The estimated per access energy for page information cache is 0.05nJ, which is consumed every time the cache is updated and read. Since the access volume varies across applications, we observe in the case of \textit{BP}, page information cache consumes significantly higher access energy than that with other applications. 

\textbf{(2) Migration:} For the migration system, we consider three data storing points as a major contributor for area, namely, NMP buffer (512B, 0.14mm$^2$), Migration queue (2KB, 0.04mm$^2$), and MDMA buffers (1KB, 0.124mm$^2$). It is worth noting that, depending on the organization and access method, the buffer and cache peripherals change significantly and so does their area. The per access energy consumption by these components are 0.122nJ, 0.02689nJ, 0.1062nJ, respectively.   

\textbf{(3) \rl~Agent:} For estimating the area and energy consumed by the \rl~agent, we focus on their major source of energy consumption that can easily be modeled as cache like structures, namely weight matrix (603KB, 2.095mm$^2$), replay buffer (36MB, 117.86mm$^2$), and state buffer (576B, 0.12mm$^2$). Their per access energy is estimated as 0.244nJ, 2.3nJ, and 0.106nJ, respectively. 

\textbf{(4) Network and Memory:} Since the migrations are realized by actually sending pages through the memory network with the help of MDMA, we also estimate the network and memory cube energy consumption by assuming 5pJ/bit/hop~\cite{poremba2017tba} and 12pJ/bit/access~\cite{pawlowski2011hmc} for the network and memory, respectively. 

\subsubsection{Overall Dynamic Energy}

In our overall dynamic energy study, we include energy consumed by (1) only additional \tech~ hardware, (2) memory network energy, and (3) memory cube energy, as major contributors to energy consumption in our framework.
In Figure~\ref{fig:energy} we observe that the energy overhead for \tech~hardware is insignificant as compared to the memory network energy consumption. There are two applications (\textit{BP} and \textit{MAC}) that increase the energy consumption by 25\% over baseline. This behavior is normal as \textit{BP} has a large number of unique pages, but its working set, the number of access per page and affinity are small/low, resulting in low page reuse and train the model with a high number of candidates, causing high dynamic energy consumption. As a solution, we can train our \rl~agent with feeding estimated energy consumption and dynamically change the training rate to achieve the performance within acceptable energy overhead, which we will explore in the future. In terms of memory network energy \tech~ observes around 20\% to 35\% increase because of the migration traffic across applications.

\section{Conclusions and Future Work}
\label{conclusion}
 Careful articulation of data placement in the physical memory cube network (MCN) becomes imperative with the advent of Near-Memory Processing (NMP) in the big data era.
 In addition, scheduling computation for both resource utilization and data co-location in large-scale NMP systems is even more challenging.
 We propose \tech, which is proven to be effective in assisting the existing NMP techniques mapped on MCN, by remapping their computation and data for improving resource utilization and optimizing communication overhead. Driven by the application's dynamic memory access behavior and the intractable size of data mapping decision space, AIMM uses Reinforcement Learning technique as an approximate solution for optimization of data and computation mapping problem. We project our technique as a plug-and-play module to be integrated with diverse NMP systems. The comprehensive experimentation shows significant performance improvement with up to 70\% speedup as compared to PEI systems for single-program workloads and up to 50\% for multi-program workloads in baseline NMP. We plan to include energy in the training process and hyper parameter tuning in the future. For broader application, \tech~can also help design data mapping in other near-data processing systems, such as processing in cache, memory, and storage.


\bibliographystyle{ieeetr}
\bibliography{ref}

\begin{thebibliography}{10}

\bibitem{thomas2014cortexsuite}
S.~Thomas, C.~Gohkale, E.~Tanuwidjaja, T.~Chong, D.~Lau, S.~Garcia, and M.~B.
  Taylor, ``Cortexsuite: A synthetic brain benchmark suite.,'' in {\em IISWC},
  pp.~76--79, 2014.

\bibitem{ahmad2015crono}
M.~Ahmad, F.~Hijaz, Q.~Shi, and O.~Khan, ``{CRONO}: {A} {B}enchmark {S}uite for
  {M}ultithreaded {G}raph {A}lgorithms {E}xecuting on {F}uturistic
  {M}ulticores,'' in {\em International Symposium on Workload Characterization
  (IISWC)}, pp.~44--55, IEEE Computer Society, 2015.

\bibitem{pawlowski2011hmc}
J.~T. Pawlowski, ``{H}ybrid {M}emory {C}ube ({HMC}),'' in {\em Hot Chips 23
  Symposium (HCS)}, pp.~1--24, IEEE, 2011.

\bibitem{lee2014hbm}
D.~U. Lee, K.~W. Kim, K.~W. Kim, H.~Kim, J.~Y. Kim, Y.~J. Park, J.~H. Kim,
  D.~S. Kim, H.~B. Park, J.~W. Shin, {\em et~al.}, ``25.2 a 1.2 v 8gb 8-channel
  128gb/s high-bandwidth memory (hbm) stacked dram with effective microbump i/o
  test methods using 29nm process and tsv,'' in {\em Solid-State Circuits
  Conference Digest of Technical Papers (ISSCC), 2014 IEEE International},
  pp.~432--433, IEEE, 2014.

\bibitem{ahn2016pei}
J.~Ahn, S.~Yoo, O.~Mutlu, and K.~Choi, ``{PIM}-{E}nabled {I}nstructions: {A}
  {L}ow-{O}verhead, {L}ocality-{A}ware {P}rocessing-in-{M}emory
  {A}rchitecture,'' in {\em International Symposium on Computer Architecture
  (ISCA)}, pp.~336--348, IEEE, 2015.

\bibitem{ahn2015tesseract}
J.~Ahn, S.~Hong, S.~Yoo, O.~Mutlu, and K.~Choi, ``{A} {S}calable
  {P}rocessing-in-{M}emory {A}ccelerator for {P}arallel {G}raph {P}rocessing,''
  in {\em International Symposium on Computer Architecture (ISCA)},
  pp.~105--117, IEEE, 2015.

\bibitem{nai2017graphpim}
L.~Nai, R.~Hadidi, J.~Sim, H.~Kim, P.~Kumar, and H.~Kim, ``{GraphPIM}:
  {E}nabling {I}nstruction-{L}evel {PIM} {O}ffloading in {G}raph {C}omputing
  {F}rameworks,'' in {\em International Symposium on High-Performance Computer
  Architecture (HPCA)}, pp.~457--468, 2017.

\bibitem{huang2019active}
J.~Huang, R.~R. Puli, P.~Majumder, S.~Kim, R.~Boyapati, K.~H. Yum, and E.~J.
  Kim, ``{Active-Routing: Compute on the Way for Near-Data Processing},'' in
  {\em 2019 IEEE International Symposium on High Performance Computer
  Architecture (HPCA)}, pp.~674--686, IEEE, 2019.

\bibitem{hsieh2016tom}
K.~Hsieh, E.~Ebrahimi, G.~Kim, N.~Chatterjee, M.~O'Connor, N.~Vijaykumar,
  O.~Mutlu, and S.~W. Keckler, ``{T}ransparent {O}ffloading and {M}apping
  ({TOM}): {E}nabling {P}rogrammer-{T}ransparent {N}ear-{D}ata {P}rocessing in
  {GPU} {S}ystems,'' in {\em International Symposium on Computer Architecture
  (ISCA)}, pp.~204--216, IEEE Press, 2016.

\bibitem{brown2020gpt3}
T.~B. Brown, B.~Mann, N.~Ryder, M.~Subbiah, J.~Kaplan, P.~Dhariwal,
  A.~Neelakantan, P.~Shyam, G.~Sastry, A.~Askell, {\em et~al.}, ``Language
  models are few-shot learners,'' {\em arXiv preprint arXiv:2005.14165}, 2020.

\bibitem{kim2013mcn}
G.~Kim, J.~Kim, J.~H. Ahn, and J.~Kim, ``{M}emory-{C}entric {S}ystem
  {I}nterconnect {D}esign with {H}ybrid {M}emory {C}ubes,'' in {\em
  International Conference on Parallel Architectures and Compilation Techniques
  (PACT)}, pp.~145--156, IEEE Press, 2013.

\bibitem{zhan2016umemnet}
J.~Zhan, I.~Akgun, J.~Zhao, A.~Davis, P.~Faraboschi, Y.~Wang, and Y.~Xie, ``{A}
  {U}nified {M}emory {N}etwork {A}rchitecture for {I}n-{M}emory {C}omputing in
  {C}ommodity {S}ervers,'' in {\em International Sympoium on Microarchitecture
  (MICRO)}, pp.~1--14, IEEE, 2016.

\bibitem{zhang2000permutation}
Z.~Zhang, Z.~Zhu, and X.~Zhang, ``A permutation-based page interleaving scheme
  to reduce row-buffer conflicts and exploit data locality,'' in {\em
  Proceedings of the 33rd annual ACM/IEEE international symposium on
  Microarchitecture}, pp.~32--41, 2000.

\bibitem{akin2015data}
B.~Akin, F.~Franchetti, and J.~C. Hoe, ``Data reorganization in memory using
  3d-stacked dram,'' {\em ACM SIGARCH Computer Architecture News}, vol.~43,
  no.~3S, pp.~131--143, 2015.

\bibitem{liu2018get}
Y.~Liu, X.~Zhao, M.~Jahre, Z.~Wang, X.~Wang, Y.~Luo, and L.~Eeckhout, ``Get out
  of the valley: power-efficient address mapping for gpus,'' in {\em 2018
  ACM/IEEE 45th Annual International Symposium on Computer Architecture
  (ISCA)}, pp.~166--179, IEEE, 2018.

\bibitem{piccoli14migration}
G.~Piccoli, H.~N. Santos, R.~E. Rodrigues, C.~Pousa, E.~Borin, and F.~M.
  Quint\~{a}o Pereira, ``Compiler support for selective page migration in numa
  architectures,'' in {\em Proceedings of the 23rd International Conference on
  Parallel Architectures and Compilation}, PACT '14, (New York, NY, USA),
  p.~369–380, Association for Computing Machinery, 2014.

\bibitem{goglin09migration}
B.~{Goglin} and N.~{Furmento}, ``Enabling high-performance memory migration for
  multithreaded applications on linux,'' in {\em 2009 IEEE International
  Symposium on Parallel Distributed Processing}, pp.~1--9, 2009.

\bibitem{chiang18numa}
M.~{Chiang}, S.~{Tu}, W.~{Su}, and C.~{Lin}, ``Enhancing inter-node process
  migration for load balancing on linux-based numa multicore systems,'' in {\em
  2018 IEEE 42nd Annual Computer Software and Applications Conference
  (COMPSAC)}, vol.~02, pp.~394--399, 2018.

\bibitem{hinuma}
Z.~{Duan}, H.~{Liu}, X.~{Liao}, H.~{Jin}, W.~{Jiang}, and Y.~{Zhang}, ``Hinuma:
  Numa-aware data placement and migration in hybrid memory systems,'' in {\em
  2019 IEEE 37th International Conference on Computer Design (ICCD)},
  pp.~367--375, 2019.

\bibitem{yazdanbakhsh2018dram}
A.~Yazdanbakhsh, C.~Song, J.~Sacks, P.~Lotfi-Kamran, H.~Esmaeilzadeh, and N.~S.
  Kim, ``In-dram near-data approximate acceleration for gpus,'' in {\em
  Proceedings of the 27th International Conference on Parallel Architectures
  and Compilation Techniques}, pp.~1--14, 2018.

\bibitem{gu2020ipim}
P.~Gu, X.~Xie, Y.~Ding, G.~Chen, W.~Zhang, D.~Niu, and Y.~Xie, ``ipim:
  Programmable in-memory image processing accelerator using near-bank
  architecture,'' in {\em 2020 ACM/IEEE 47th Annual International Symposium on
  Computer Architecture (ISCA)}, pp.~804--817, IEEE, 2020.

\bibitem{kwon202125}
Y.-C. Kwon, S.~H. Lee, J.~Lee, S.-H. Kwon, J.~M. Ryu, J.-P. Son, O.~Seongil,
  H.-S. Yu, H.~Lee, S.~Y. Kim, {\em et~al.}, ``25.4 a 20nm 6gb
  function-in-memory dram, based on hbm2 with a 1.2 tflops programmable
  computing unit using bank-level parallelism, for machine learning
  applications,'' in {\em 2021 IEEE International Solid-State Circuits
  Conference (ISSCC)}, vol.~64, pp.~350--352, IEEE, 2021.

\bibitem{alian2019netdimm}
M.~Alian and N.~S. Kim, ``Netdimm: Low-latency near-memory network interface
  architecture,'' in {\em Proceedings of the 52nd Annual IEEE/ACM International
  Symposium on Microarchitecture}, pp.~699--711, 2019.

\bibitem{alian2018application}
M.~Alian, S.~W. Min, H.~Asgharimoghaddam, A.~Dhar, D.~K. Wang, T.~Roewer,
  A.~McPadden, O.~O'Halloran, D.~Chen, J.~Xiong, {\em et~al.},
  ``Application-transparent near-memory processing architecture with memory
  channel network,'' in {\em 2018 51st Annual IEEE/ACM International Symposium
  on Microarchitecture (MICRO)}, pp.~802--814, IEEE, 2018.

\bibitem{huangfu2019medal}
W.~Huangfu, X.~Li, S.~Li, X.~Hu, P.~Gu, and Y.~Xie, ``Medal: Scalable dimm
  based near data processing accelerator for dna seeding algorithm,'' in {\em
  Proceedings of the 52nd Annual IEEE/ACM International Symposium on
  Microarchitecture}, pp.~587--599, 2019.

\bibitem{kwon2019tensordimm}
Y.~Kwon, Y.~Lee, and M.~Rhu, ``Tensordimm: A practical near-memory processing
  architecture for embeddings and tensor operations in deep learning,'' in {\em
  Proceedings of the 52nd Annual IEEE/ACM International Symposium on
  Microarchitecture}, pp.~740--753, 2019.

\bibitem{ke2020recnmp}
L.~Ke, U.~Gupta, B.~Y. Cho, D.~Brooks, V.~Chandra, U.~Diril, A.~Firoozshahian,
  K.~Hazelwood, B.~Jia, H.-H.~S. Lee, {\em et~al.}, ``Recnmp: Accelerating
  personalized recommendation with near-memory processing,'' in {\em 2020
  ACM/IEEE 47th Annual International Symposium on Computer Architecture
  (ISCA)}, pp.~790--803, IEEE, 2020.

\bibitem{tanenbaum-osbook}
A.~S. Tanenbaum and H.~Bos, {\em Modern Operating Systems}.
\newblock USA: Prentice Hall Press, 4th~ed., 2014.

\bibitem{hoard}
E.~D. Berger, K.~S. McKinley, R.~D. Blumofe, and P.~R. Wilson, ``Hoard: A
  scalable memory allocator for multithreaded applications,'' {\em SIGPLAN
  Not.}, vol.~35, p.~117–128, Nov. 2000.

\bibitem{locality_of_ref}
R.~Courts, ``Improving locality of reference in a garbage-collecting memory
  management system,'' {\em Commun. ACM}, vol.~31, p.~1128–1138, Sept. 1988.

\bibitem{dynamic-storage-allocation}
P.~R. Wilson, M.~S. Johnstone, M.~Neely, and D.~Boles, ``Dynamic storage
  allocation: A survey and critical review,'' in {\em Memory Management} (H.~G.
  Baler, ed.), (Berlin, Heidelberg), pp.~1--116, Springer Berlin Heidelberg,
  1995.

\bibitem{locality-improving_DMA}
Y.~Feng and E.~D. Berger, ``A locality-improving dynamic memory allocator,'' in
  {\em Proceedings of the 2005 Workshop on Memory System Performance}, MSP '05,
  (New York, NY, USA), p.~68–77, Association for Computing Machinery, 2005.

\bibitem{buddy}
K.~C. Knowlton, ``A fast storage allocator,'' {\em Commun. ACM}, vol.~8,
  p.~623–624, Oct. 1965.

\bibitem{lkml-transparent-huge-pages}
J.~Corbet, ``{LKML}: Transparent huge pages in 2.6.38,'' Jan. 2011.

\bibitem{hawkeye}
A.~Panwar, S.~Bansal, and K.~Gopinath, ``Hawkeye: Efficient fine-grained os
  support for huge pages,'' in {\em Proceedings of the Twenty-Fourth
  International Conference on Architectural Support for Programming Languages
  and Operating Systems}, ASPLOS '19, (New York, NY, USA), p.~347–360,
  Association for Computing Machinery, 2019.

\bibitem{lisp-gc}
L.~P. Deutsch and D.~G. Bobrow, ``An efficient, incremental, automatic garbage
  collector,'' {\em Commun. ACM}, vol.~19, p.~522–526, Sept. 1976.

\bibitem{lieberman-gc-83}
H.~Lieberman and C.~Hewitt, ``A real-time garbage collector based on the
  lifetimes of objects,'' {\em Commun. ACM}, vol.~26, p.~419–429, June 1983.

\bibitem{generation-gc}
A.~W. Appel, ``Simple generational garbage collection and fast allocation,''
  {\em Softw. Pract. Exper.}, vol.~19, p.~171–183, Feb. 1989.

\bibitem{java-gc}
``{O}racle {JDK} 9 documentation - 5 available collectors, howpublished =
  {\url{https://docs.oracle.com/javase/9/gctuning/available-collectors.htm#jsgct-guid-f215a508-9e58-40b4-90a5-74e29bf3bd3c}},,''

\bibitem{bolosky1991numa}
W.~J. Bolosky, M.~L. Scott, R.~P. Fitzgerald, R.~J. Fowler, and A.~L. Cox,
  ``Numa policies and their relation to memory architecture,'' {\em ACM SIGOPS
  Operating Systems Review}, vol.~25, no.~Special Issue, pp.~212--221, 1991.

\bibitem{nikolopoulos00migration}
D.~S. {Nikolopoulos}, T.~S. {Papatheodorou}, C.~D. {Polychronopoulos},
  J.~{Labarta}, and E.~{Ayguade}, ``User-level dynamic page migration for
  multiprogrammed shared-memory multiprocessors,'' in {\em Proceedings 2000
  International Conference on Parallel Processing}, pp.~95--103, 2000.

\bibitem{yao81mem-alloc}
A.~C. Yao, ``An analysis of a memory allocation scheme for implementing
  stacks,'' {\em SIAM Journal on Computing}, vol.~10, no.~2, pp.~398--403,
  1981.

\bibitem{ribeiro09affinity}
C.~P. {Ribeiro}, J.~{Mehaut}, A.~{Carissimi}, M.~{Castro}, and L.~G.
  {Fernandes}, ``Memory affinity for hierarchical shared memory
  multiprocessors,'' in {\em 2009 21st International Symposium on Computer
  Architecture and High Performance Computing}, pp.~59--66, 2009.

\bibitem{lameter13numa}
C.~Lameter, ``Numa (non-uniform memory access): An overview,'' {\em Queue},
  vol.~11, 07 2013.

\bibitem{baruah2020griffin}
T.~Baruah, Y.~Sun, A.~T. Din{\c{c}}er, S.~A. Mojumder, J.~L. Abell{\'a}n,
  Y.~Ukidave, A.~Joshi, N.~Rubin, J.~Kim, and D.~Kaeli, ``Griffin:
  Hardware-software support for efficient page migration in multi-gpu
  systems,'' in {\em 2020 IEEE International Symposium on High Performance
  Computer Architecture (HPCA)}, pp.~596--609, IEEE, 2020.

\bibitem{sutton2018reinforcement}
R.~S. Sutton and A.~G. Barto, {\em Reinforcement learning: An introduction}.
\newblock MIT press, 2018.

\bibitem{mnih2015human}
V.~Mnih, K.~Kavukcuoglu, D.~Silver, A.~A. Rusu, J.~Veness, M.~G. Bellemare,
  A.~Graves, M.~Riedmiller, A.~K. Fidjeland, G.~Ostrovski, {\em et~al.},
  ``Human-level control through deep reinforcement learning,'' {\em nature},
  vol.~518, no.~7540, pp.~529--533, 2015.

\bibitem{won2014up}
J.-Y. Won, X.~Chen, P.~Gratz, J.~Hu, and V.~Soteriou, ``Up by their bootstraps:
  Online learning in artificial neural networks for cmp uncore power
  management,'' in {\em 2014 IEEE 20th International Symposium on High
  Performance Computer Architecture (HPCA)}, pp.~308--319, IEEE, 2014.

\bibitem{lin2020deep}
T.-R. Lin, D.~Penney, M.~Pedram, and L.~Chen, ``A deep reinforcement learning
  framework for architectural exploration: A routerless noc case study,'' in
  {\em 2020 IEEE International Symposium on High Performance Computer
  Architecture (HPCA)}, pp.~99--110, IEEE, 2020.

\bibitem{yin2020experiences}
J.~Yin, S.~Sethumurugan, Y.~Eckert, C.~Patel, A.~Smith, E.~Morton, M.~Oskin,
  N.~E. Jerger, and G.~H. Loh, ``Experiences with ml-driven design: A noc case
  study,'' in {\em 2020 IEEE International Symposium on High Performance
  Computer Architecture (HPCA)}, pp.~637--648, IEEE, 2020.

\bibitem{wang2019intellinoc}
K.~Wang, A.~Louri, A.~Karanth, and R.~Bunescu, ``Intellinoc: A holistic design
  framework for energy-efficient and reliable on-chip communication for
  manycores,'' in {\em 2019 ACM/IEEE 46th Annual International Symposium on
  Computer Architecture (ISCA)}, pp.~1--12, IEEE, 2019.

\bibitem{peled2015semantic}
L.~Peled, S.~Mannor, U.~Weiser, and Y.~Etsion, ``Semantic locality and
  context-based prefetching using reinforcement learning,'' in {\em 2015
  ACM/IEEE 42nd Annual International Symposium on Computer Architecture
  (ISCA)}, pp.~285--297, IEEE, 2015.

\bibitem{ipek2008self}
E.~Ipek, O.~Mutlu, J.~F. Mart{\'\i}nez, and R.~Caruana, ``Self-optimizing
  memory controllers: A reinforcement learning approach,'' {\em ACM SIGARCH
  Computer Architecture News}, vol.~36, no.~3, pp.~39--50, 2008.

\bibitem{ahn2019reinforcement}
B.~H. Ahn, P.~Pilligundla, and H.~Esmaeilzadeh, ``Reinforcement learning and
  adaptive sampling for optimized dnn compilation,'' {\em arXiv preprint
  arXiv:1905.12799}, 2019.

\bibitem{kao2020confuciux}
S.-C. Kao, G.~Jeong, and T.~Krishna, ``Confuciux: Autonomous hardware resource
  assignment for dnn accelerators using reinforcement learning,'' in {\em 2020
  53rd Annual IEEE/ACM International Symposium on Microarchitecture (MICRO)},
  pp.~622--636, IEEE, 2020.

\bibitem{wu2020core}
N.~Wu, L.~Deng, G.~Li, and Y.~Xie, ``Core placement optimization for multi-chip
  many-core neural network systems with reinforcement learning,'' {\em ACM
  Transactions on Design Automation of Electronic Systems (TODAES)}, vol.~26,
  no.~2, pp.~1--27, 2020.

\bibitem{fa3c}
H.~Cho, P.~Oh, J.~Park, W.~Jung, and J.~Lee, ``Fa3c: Fpga-accelerated deep
  reinforcement learning,'' in {\em Proceedings of the Twenty-Fourth
  International Conference on Architectural Support for Programming Languages
  and Operating Systems}, ASPLOS '19, (New York, NY, USA), p.~499–513,
  Association for Computing Machinery, 2019.

\bibitem{deepqaccelerator}
J.~Su, J.~Liu, D.~B. Thomas, and P.~Y. Cheung, ``Neural network based
  reinforcement learning acceleration on fpga platforms,'' {\em SIGARCH Comput.
  Archit. News}, vol.~44, p.~68–73, Jan. 2017.

\bibitem{plappert2016kerasrl}
M.~Plappert, ``keras-rl.'' \url{https://github.com/keras-rl/keras-rl}, 2016.

\bibitem{brockman2016openai}
G.~Brockman, V.~Cheung, L.~Pettersson, J.~Schneider, J.~Schulman, J.~Tang, and
  W.~Zaremba, ``Openai gym,'' {\em arXiv preprint arXiv:1606.01540}, 2016.

\bibitem{balasubramonian2017cacti}
R.~Balasubramonian, A.~B. Kahng, N.~Muralimanohar, A.~Shafiee, and V.~Srinivas,
  ``Cacti 7: New tools for interconnect exploration in innovative off-chip
  memories,'' {\em ACM Transactions on Architecture and Code Optimization
  (TACO)}, vol.~14, no.~2, pp.~1--25, 2017.

\bibitem{ptmalloc}
D.~Delorie, ``Glibc wiki - overview of malloc.''
  \url{https://sourceware.org/glibc/wiki/MallocInternals}.

\bibitem{tcmalloc}
``{TCMalloc : Thread-Caching Malloc}.''
  \url{https://google.github.io/tcmalloc/design.html}.

\bibitem{jemalloc}
``jemalloc.'' \url{http://jemalloc.net/}.

\bibitem{che2010characterization}
S.~Che, J.~W. Sheaffer, M.~Boyer, L.~G. Szafaryn, L.~Wang, and K.~Skadron, ``A
  characterization of the rodinia benchmark suite with comparison to
  contemporary cmp workloads,'' in {\em IEEE International Symposium on
  Workload Characterization (IISWC'10)}, pp.~1--11, IEEE, 2010.

\bibitem{che2009rodinia}
S.~Che, M.~Boyer, J.~Meng, D.~Tarjan, J.~W. Sheaffer, S.-H. Lee, and
  K.~Skadron, ``{R}odinia: {A} {B}enchmark {S}uite for {H}eterogeneous
  {C}omputing,'' in {\em International Symposium on Workload Characterization
  (IISWC)}, pp.~44--54, IEEE Computer Society, 2009.

\bibitem{hinton2007boltzmann}
G.~E. Hinton, ``Boltzmann machine,'' {\em Scholarpedia}, vol.~2, no.~5,
  p.~1668, 2007.

\bibitem{bienia2008parsec}
C.~Bienia, S.~Kumar, J.~P. Singh, and K.~Li, ``The parsec benchmark suite:
  Characterization and architectural implications,'' in {\em Proceedings of the
  17th International Conference on Parallel Architectures and Compilation
  Techniques}, PACT '08, (New York, NY, USA), pp.~72--81, ACM, 2008.

\bibitem{poremba2017tba}
M.~Poremba, I.~Akgun, J.~Yin, O.~Kayiran, Y.~Xie, and G.~H. Loh, ``{T}here and
  {B}ack {A}gain: {O}ptimizing the {I}nterconnect in {N}etworks of {M}emory
  {C}ubes,'' in {\em International Symposium on Computer Architecture (ISCA)},
  pp.~678--690, ACM, 2017.

\end{thebibliography}

\end{document}